\documentclass[11pt]{article}

\usepackage{moriond}
\usepackage{graphicx}
\begin{document}
\vspace*{4cm}
\title{Positioning and orienting a static cylindrical radio-reflector
  for wide field surveys}

\author{Moniez, Marc}

\address{Laboratoire de l'Acc\'{e}l\'{e}rateur Lin\'{e}aire,
{\sc IN2P3-CNRS}, Universit\'e de Paris-Sud, \\
B.P. 34, 91898 Orsay Cedex, France.
E-mail: moniez@lal.in2p3.fr}

\maketitle\abstracts{
Several projects in radioastronomy plan to use large static cylindrical
reflectors with an extended lobe sampling a sector of the rotating sky.
This study provides the exact mathematical expression of the transit
time of a celestial object within the acceptance lobe of such a
cylindrical device. The mathematical approach, based on the
stereographic projection, allows one to study the optimisation of the position and
orientation of the radio-reflector, and should provide exact coefficients
for the spatial Fourier Transform of the radio signal along the
cylinder axis. 
}
\section{Introduction}
\begin{figure}[h]
\parbox{7cm}{
\includegraphics[width=7cm]{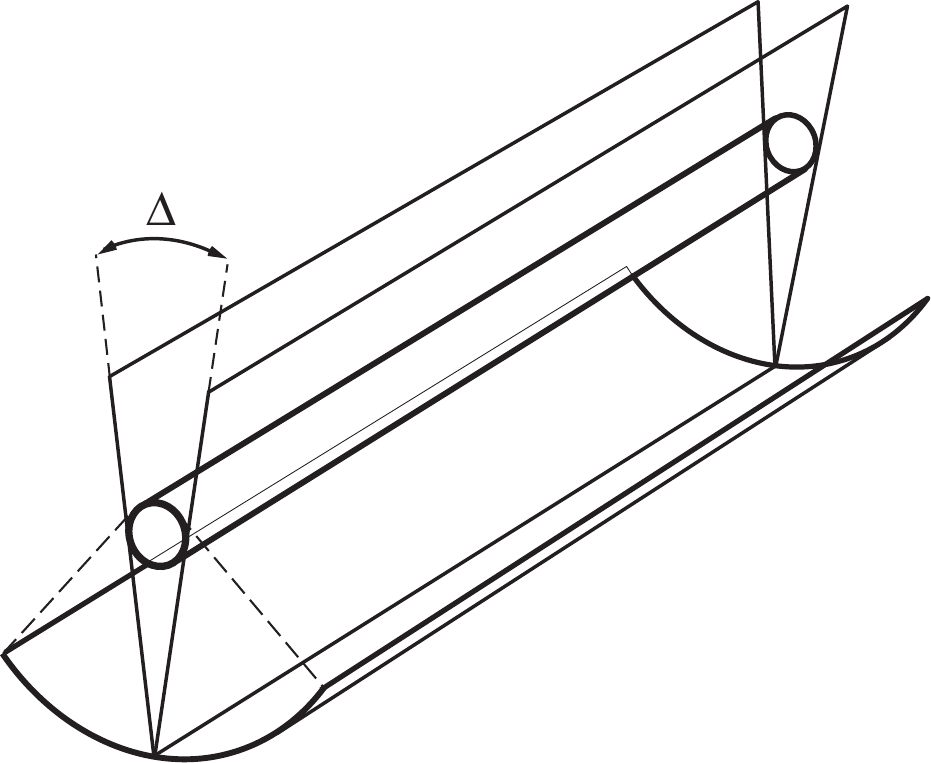}
}
\hspace{0.5cm}
\parbox{6cm}{
\caption[]{\it The reflector and its field of view,
defined as the angular sector $\Delta$ that is focalised
within the antenna's acceptance.
}
}
\label{secteur}
\end{figure}

Several baryonic oscillation (BAO) radio projects \cite{1,2,3}
plan to operate a series of
parallel static reflectors of large parabolic cylinder shape,
to map the $21 cm$ HI emission line.
The sky will transit over the acceptance lobe, which is
defined by an angular sector of aperture $\Delta$ centered around a vertical
plane (see Fig. \ref{secteur}).

\begin{figure}[h]
\centering
\vbox{
\includegraphics[width=12cm]{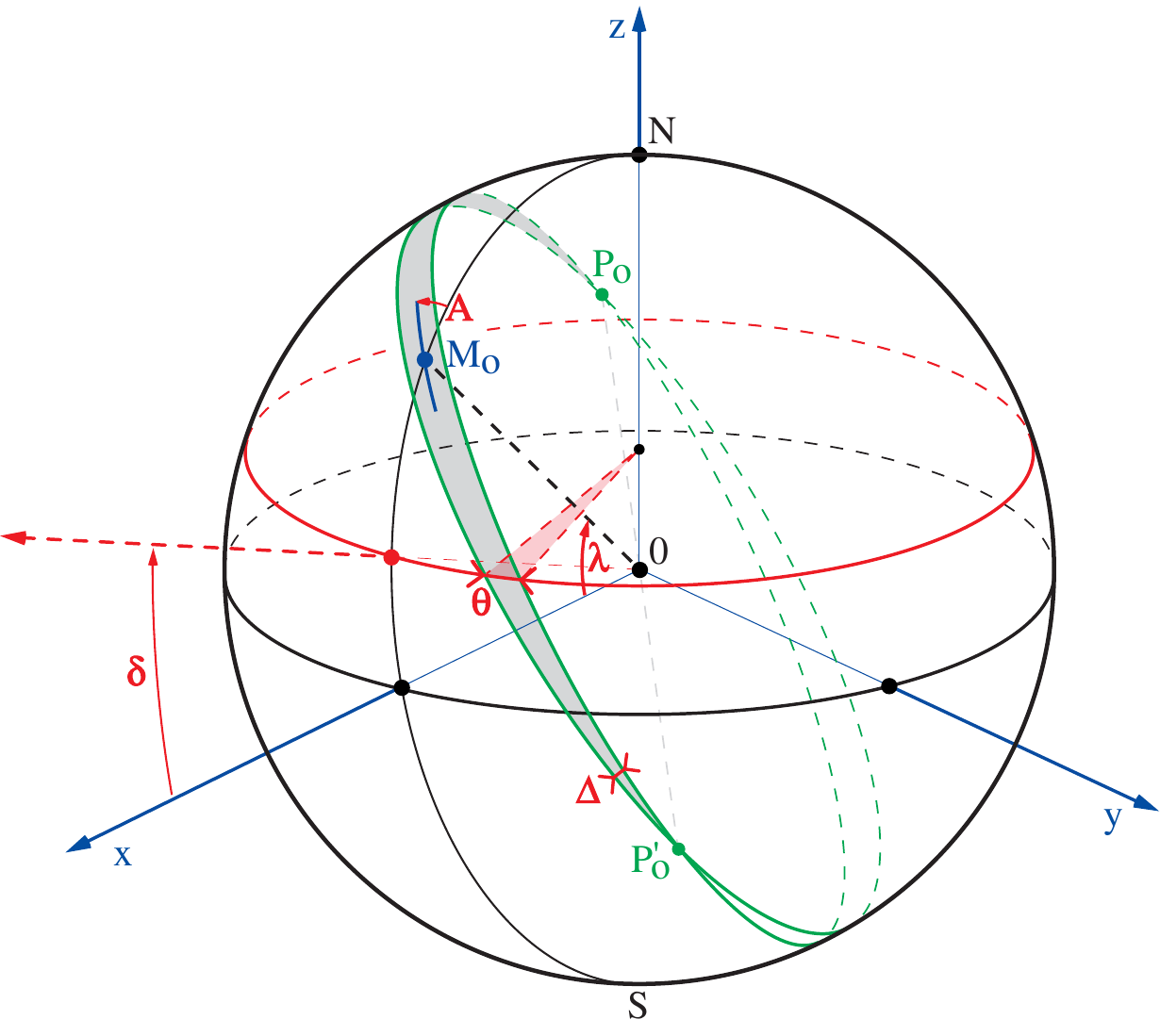}
}
\caption[]{\it The celestial sphere with the projected position
of the observer $M_0$ (latitude $\lambda$), the projected
orientation of the reflector (A) and the projected portion of
detectable sky (detection lobe), defined as the angular sector
$\Delta$ of axis $P_0 P'_0$ (in grey),
where $P_0$ and $P'_0$ are the projections of the reflector's
axis. $\theta/2\pi$ is the fraction of the sideral day that an object
of declination $\delta$ will spend within the detection lobe.
}
\label{vision3D}
\end{figure}

The projection on the sky of this
angular sector can be seen on Fig. \ref{vision3D}.
In this paper, I produce the exact calculation of the transit
time of a celestial object, as a function of its declination.
In section 5, I use the results of the calculation to compare
the performances of various radio-telescope latitudes
(France, Morocco, South Africa, equator) and configurations (orientation).
In particular, we show that the North-South orientation usually considered
may not be the optimal one for high-z BAO studies that
need large exposure times, but not necessarilly the largest possible
field of view.

Another possible use of the exact expression for the transit time is the
production of exact coefficients for Fourier Transform calculations
along the cylinder axis.

\section{Notations}
We will use the following notations (see Fig. \ref{vision3D}):
\begin{itemize}
\item
$\lambda$ is the observatory's latitude,
\item
$M_0$ its position on Earth.
\item
$A$ is the azimuth of the reflector (with respect to the meridian).
\item
$\Delta$ is the lobe's aperture. A celestial object can be detected
only if it enters this lobe.
\item
$P_0$ and $P'_0$ are the intersections of the lobe's definition planes
on the celestial sphere. $P_0P'_0M_0$ define a large circle
on the sphere, with $(\widehat{P_0OM_0})=(\widehat{M_0OP'_0})=\pi/2$.
\item
$\delta$ is the declination of a celestial object.
\end{itemize}
The (sideral) daily exposure of an object is given by the fraction of its corresponding
parallel that is included in the acceptance lobe. 
On Fig. \ref{vision3D}, this exposure is given by $\theta/2\pi \times
1\, sideral\, day$.
When $\lambda$, $A$ and $\Delta$ are defined,
it depends only on the declination of the object.
From the figure, it can be seen that the daily exposure is in general not
uniform for a random choice of $\lambda$ and $A$. The objective of this
paper is to systematically study the exposure as a function of the declination
for any antenna configuration, and to provide an optimization tool.

\section{The stereographic projection}
\begin{figure}[h]
\centering
\parbox{8cm}{
\includegraphics[width=8cm]{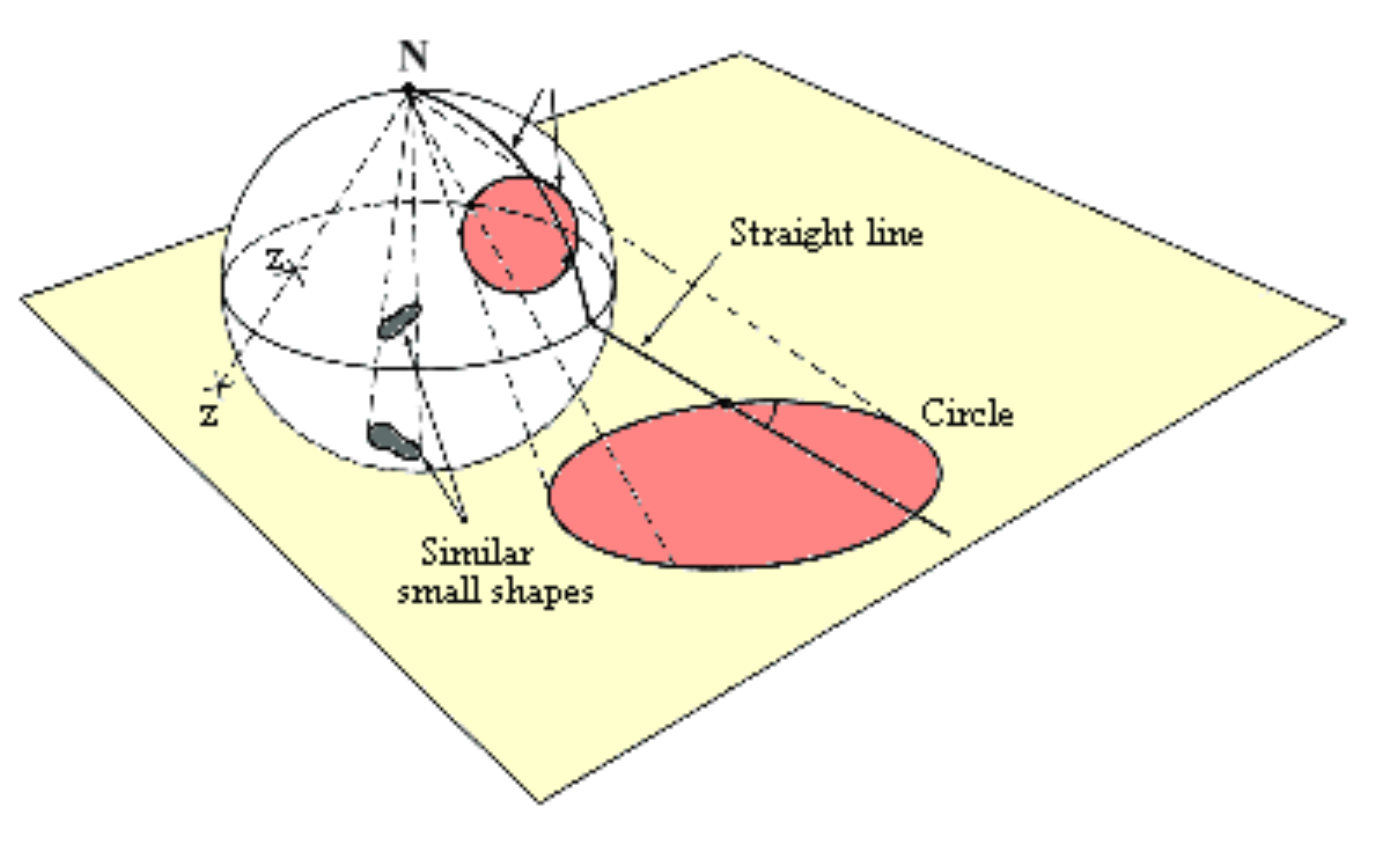}
}
\hspace{0.5cm}
\parbox{6cm}{
\caption[]{\it Stereographic projection from the North pole.
}
}
\label{stereographic}
\end{figure}
\begin{figure}[h]
\centering
\vbox{
\includegraphics[width=13cm]{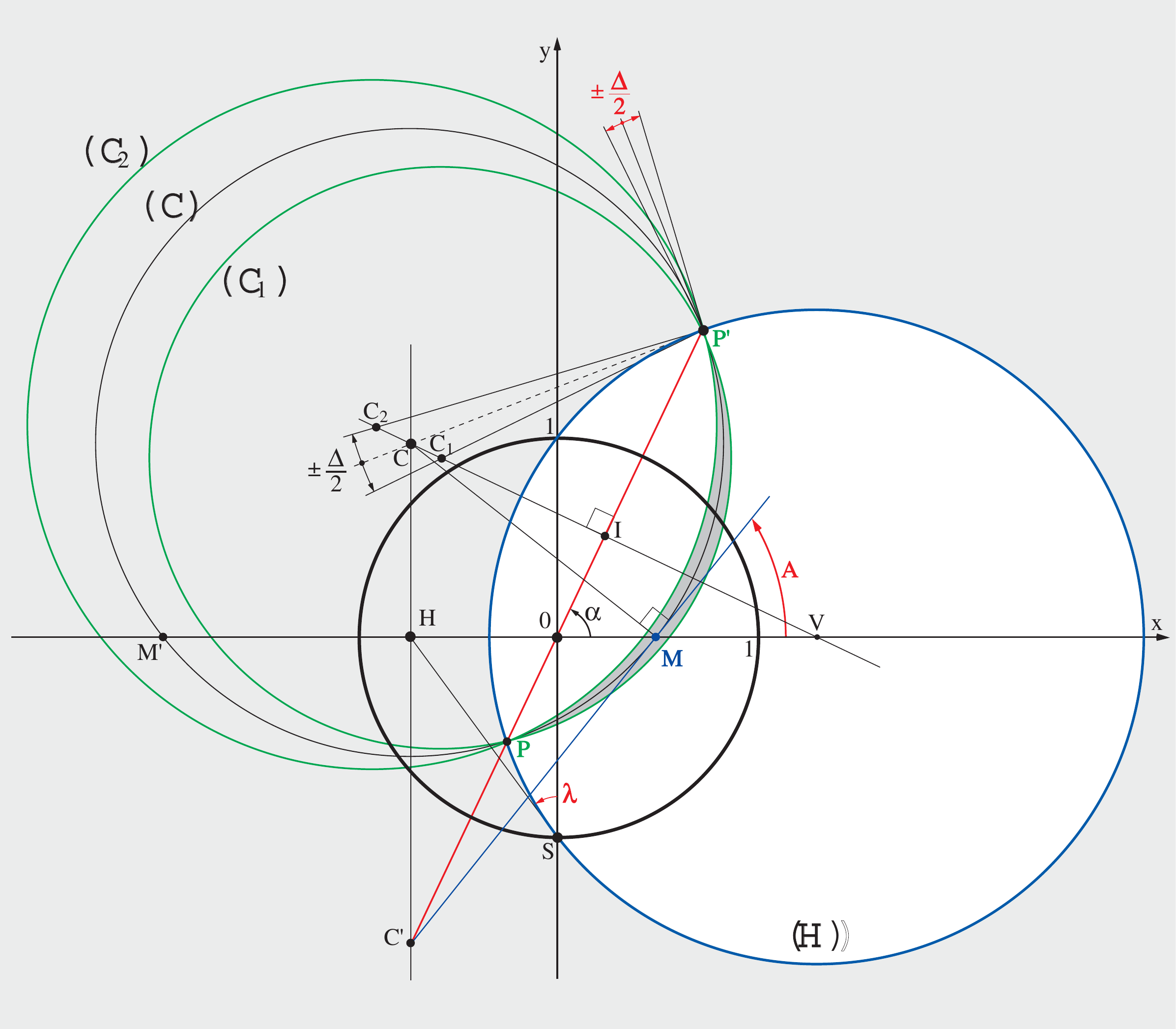}
}
\caption[]{
\it The stereographic projection of Fig. \ref{vision3D}.
This construction is drawn from point $S$ as follows:

- $H$ is such that $(\widehat{OSH})=\lambda$.

- $M$ and $M'$, projections of $M_0$ and its antipod $M'_0$ are
the intersections of $Ox$ with the circle
of radius $HS$, centered on $H$.

- $C'$ is the intersection of the vertical line containing $H$
and the line defined by $M$ (or $M'$) and angle $A$ (or $-A$).

- $C$ is the intersection of the vertical line containing $H$
and the line defined by $M$ (or $M'$) and angle $A+\pi/2$
(or $-A+\pi/2$).

- $P$ and $P'$ are the intersections of the line $OC'$ and the
circle ${\cal C}$ of radius $CM$ centered on $C$.

- $I$ is the center of $P P'$.

- The horizon circle (${\cal H}$) is centered on V, intersection
of $0x$ with $IC$.
}
\label{proj2D}
\end{figure}

The geometrical tool used to establish the exposure versus declination
function is the stereographic projection, because of the geometrical
configuration includes
only circles, and mainly large circles (see Fig. \ref{stereographic}).
Fig. \ref{vision3D} is then
projected from the South pole on the equatorial plane (Fig. \ref{proj2D}).

The main properties of the stereographic projection that we will use
are the following:
\begin{itemize}
\item
The projection of a circle on the sphere is a circle or a straight line
on the plane.
\item
The projection of a large circle is a circle (or a straight line)
that intercepts the equator in 2 diametrally opposite points.
\item
The projection of a meridian is a straight line that includes the origin.
\item
Angles between tangents on the sphere are invariant under the projection.
\item
Lengths and surfaces are not invariant under the projection, but for
symmetry reasons, the scaling is constant along a given parallel.
The fraction of a parallel that is included in the acceptance lobe will
then be invariant under the projection.
\end{itemize}
On figure \ref{proj2D}, the thick circle (of radius 1)
is the equator, the crescent is the projection of the lobe, and
the blue circle (${\cal H}$) is the projection of the horizon of $M_0$
(projected on $M$).
We do not restrict the generality by assuming that the observatory is
located in the northern hemisphere (in the reverse case, juste exchange
North and South) and that $0<A<\pi/2$.
The projection of the visible part of the sky is then given by the white
(not shaded) area (that contains $M$ ($0<x_M<1$), projection of $M_0$).
The two circles (${\cal C}_1$) and (${\cal C}_2$) are the projections
of the large circles defining the lobe, that intercept each other at $P_0$ and $P'_0$.
Since $P_0$ and $P'_0$ are on the same meridian (because they are antipodic),
their projections $P$ and $P'$ are aligned with the origin.
As a consequence of the angle conservation, the
projection of the large circle defining the
median plane of the lobe intercepts the observer's meridian
projection (Ox axis) at angle $A$.
We define $I$ as the center of $PP'$ segment.

The horizon circle (${\cal H}$), projection of the large circle
horizon of $M_0$, contains $P$, $P'$, and intersects the equator at
$S(0,-1)$ and $(0,1)$ with angle $\pi/2-\lambda$.
Its center $V$ is aligned with $IC$ which is the median of $PP'$
\footnote{This circle is described by $P$ and $P'$ when the angle
$A$ varies.} for symmetry reasons.
\subsection{Useful relations}
Our aim is to find the intersections of the projected lobe sides
(${\cal C}_1$) and (${\cal C}_2$) with a
given declination circle.
In this subsection, we first determine $OI$, $IP$
and $IC$ that are needed to establish the equations of these circles.\\
Fig. 5
shows some important geometrical relations
involving the observer's position and its antipodic point,
in the transverse view of the stereographic projection.
\begin{figure}[h]
\parbox{7.5cm}{
\includegraphics[width=7.5cm]{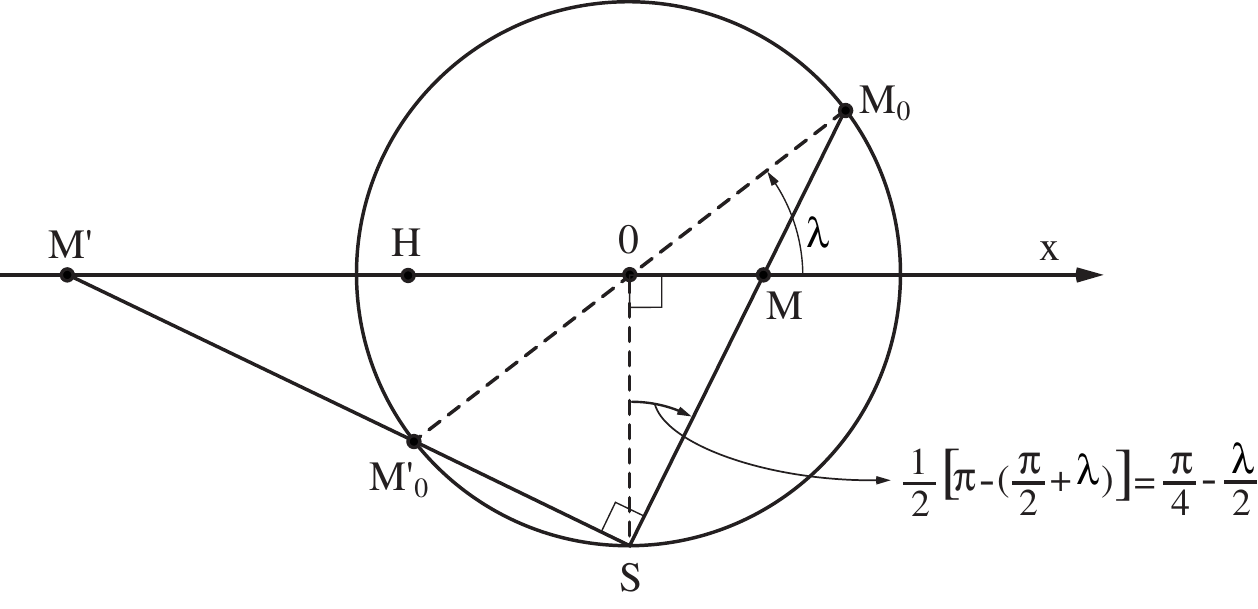}
}
\hspace{0.3cm}
\parbox{8cm}{
\caption[]{\it Vertical view of the stereographic sphere
along the X axis, showing the relations between the projections
$M$ and $M'$ of two antipodic points $M_0$ (at latitude $\lambda$)
and $M'_0$. $S$ is the south pole.
\\
$OM^2+OS^2=MS^2$
\\
$OM'^2+OS^2=M'S^2$
\\
$MS^2+M'S^2=MM'^2$
\\
The two first relations reported in the third one give
\\
$MM'^2=OM^2+OM'^2+2 OS^2$
\\
which is also equal to
\\
$MM'^2=(MO+OM')^2=OM^2+OM'^2+2.OM.OM'$
\\
It follows that $OM\times OM'=OS^2=1$.
}
}
\label{trirect}
\end{figure}

These relations are valid for any
couple of antipodic points.
The following series of relations allows to extract the most
pertinent parameters of the lobe projection for subsequent calculation
of the exposure time.
\begin{itemize}
\item
\begin{equation}
OS=1
\end{equation}
\item
Let $H$ be the center of $MM'$ where $M'$ is the projection of $M'_0$,
antipodic of $M_0$ :
\begin{equation}
OH=\tan \lambda
\label{OH}
\end{equation}
because $H$ is the center of the circle circumscribing the $MSM'$
right triangle (Fig. 5), implying that $HS=HM$ and therefore
$(\widehat{HMS})=(\widehat{MSH})=\pi/4+\lambda/2$
and $(\widehat{OSH})=(\widehat{MSH})-(\widehat{MSO})=\lambda$.
\item
It also follows from this that
\begin{equation}
HM=HS=1/\cos\lambda
\label{HM}
\end{equation}
\item
As the horizon circle (${\cal H}$) intersects the equator
with angle $\pi/2-\lambda$, it is tangent to $HS$;
then, taking into account that $OS=1$, its center $V$ is such that
\begin{equation}
OV=\cot\lambda
\label{OV}
\end{equation}
and its radius is
\begin{equation}
VS=1/\sin\lambda.
\end{equation}
\item
Let $C$ be the center of circle (${\cal C}$)  projected from the lobes'
median large circle.\\
i) (${\cal C}$) includes $M$, $M'$, $P$ and $P'$, then its center $C$
belongs to the median of $MM'$ (defined by $HC$ on Fig. \ref{proj2D}).\\
ii) The angle between $Ox$ (projection of the meridian) and the tangent of
(${\cal C}$) at $M$ is $A$, by virtue of the angle conservation.
$CM$ is then orthogonal to that tangent (see Fig. \ref{proj2D}).
It follows that :
\begin{equation}
HC=\frac{HM}{\tan A} =\frac{1}{\cos\lambda \tan A}.
\label{HC}
\end{equation}
\item
We also define $C'$ (bottom-left in Fig. \ref{proj2D})
as the center of the circle projected from the large
circle perpendicular to the lobe at $M_0$.\\
i) This projected circle includes $M$ and $M'$, then $C'$ also
belongs to the median of $MM'$ defined by $HC$.\\
ii) Since this large circle is orthogonal to the lobe,
its projected circle is orthogonal to (${\cal C}$) at $M$ ;
it follows that $C'M$ is tangent to (${\cal C}$) at $M$
(see Fig. \ref{proj2D}).\\
iii) Since $P_0$ and $P_0'$ are the poles of this large circle, for
symmetry reasons, this circle intersects the $P_0P_0'$ meridian
with right angle. It follows that $PP'$ (aligned with $O$
because $P_0$ and $P_0'$ are on the same meridian) is perpendicular
to the projected circle, and subsequently aligned with its center $C'$
(see Fig. \ref{proj2D}).\\
It follows that :
\begin{equation}
HC'=HM.\tan A =\frac{\tan A}{\cos\lambda}.
\label{HC'}
\end{equation}
\item
\begin{equation}
CC'=HC+HC'=\frac{1}{\cos\lambda \tan A}+\frac{\tan A}{\cos\lambda}
=\frac{1}{\cos\lambda \cos A\sin A}.
\label{CC'}
\end{equation}
\item
The angle $\alpha$, as marked on Fig. \ref{proj2D} will be useful for
subsequent calculations.
\begin{equation}
HC'=OH\tan \alpha =\tan \lambda \tan \alpha \ (using (\ref{OH})).
\label{toto1}
\end{equation}
Combining with (\ref{HC'}), one obtains
\begin{equation}
\tan A =\sin \lambda \tan \alpha.
\label{tgalpha}
\end{equation}
\item
$CC'^2=CI^2+C'I^2$ and $C'I/CI=\tan \alpha =\tan A /\sin \lambda$ and
(\ref{CC'}) $=>$
\begin{equation}
IC\sqrt{1+\tan^2 A/\sin^2 \lambda}=\frac{1}{\cos\lambda \cos A \sin A}.
\label{toto2}
\end{equation}
Then
\begin{equation}
IC=\frac{\tan \lambda}{\sin A\cos A}\frac{1}{\sqrt{sin^2\lambda+\tan^2 A}}
\label{IC}
\end{equation}
and
\begin{equation}
IC'=IC\tan \alpha =IC\tan A /\sin \lambda =
\frac{1}{\cos\lambda \cos^2 A }\frac{1}{\sqrt{\sin^2 \lambda+\tan^2 A}}.
\label{IC'}
\end{equation}
\item
Using relations (\ref{OH}) and (\ref{HC'}) one finds:
\begin{equation}
OC'=\sqrt{OH^2+HC'^2}=\sqrt{\tan^2 \lambda +\frac{\tan^2 A}{\cos^2 \lambda }}
=\frac{\sqrt{\sin^2 \lambda +\tan^2 A}}{\cos\lambda}.
\label{OC'}
\end{equation}
\item
$OI=IC'-OC'$.
Using (\ref{IC'}) and (\ref{OC'}), one obtains, after simplification:
\begin{equation}
OI=\frac{\cos\lambda}{\sqrt{\sin^2 \lambda+\tan^2 A}}.
\label{OI}
\end{equation}
\item
$P$ and $P'$ are the images of antipodic points, equivalently to
$M$ and $M'$. Using
Fig. 5
the relation:
$OH^2+OS^2=HS^2=HM^2$ can be transposed as $OI^2+1=IP^2$.
Then
\begin{equation}
IP=\sqrt{1+OI^2}=\frac{1}{\cos A }\frac{1}{\sqrt{\sin^2 \lambda +\tan^2 A}}.
\label{IP}
\end{equation}
\end{itemize}

\subsection{Expression of the exposure time}
\begin{figure}[h]
\centering
\vbox{
\includegraphics[width=14cm]{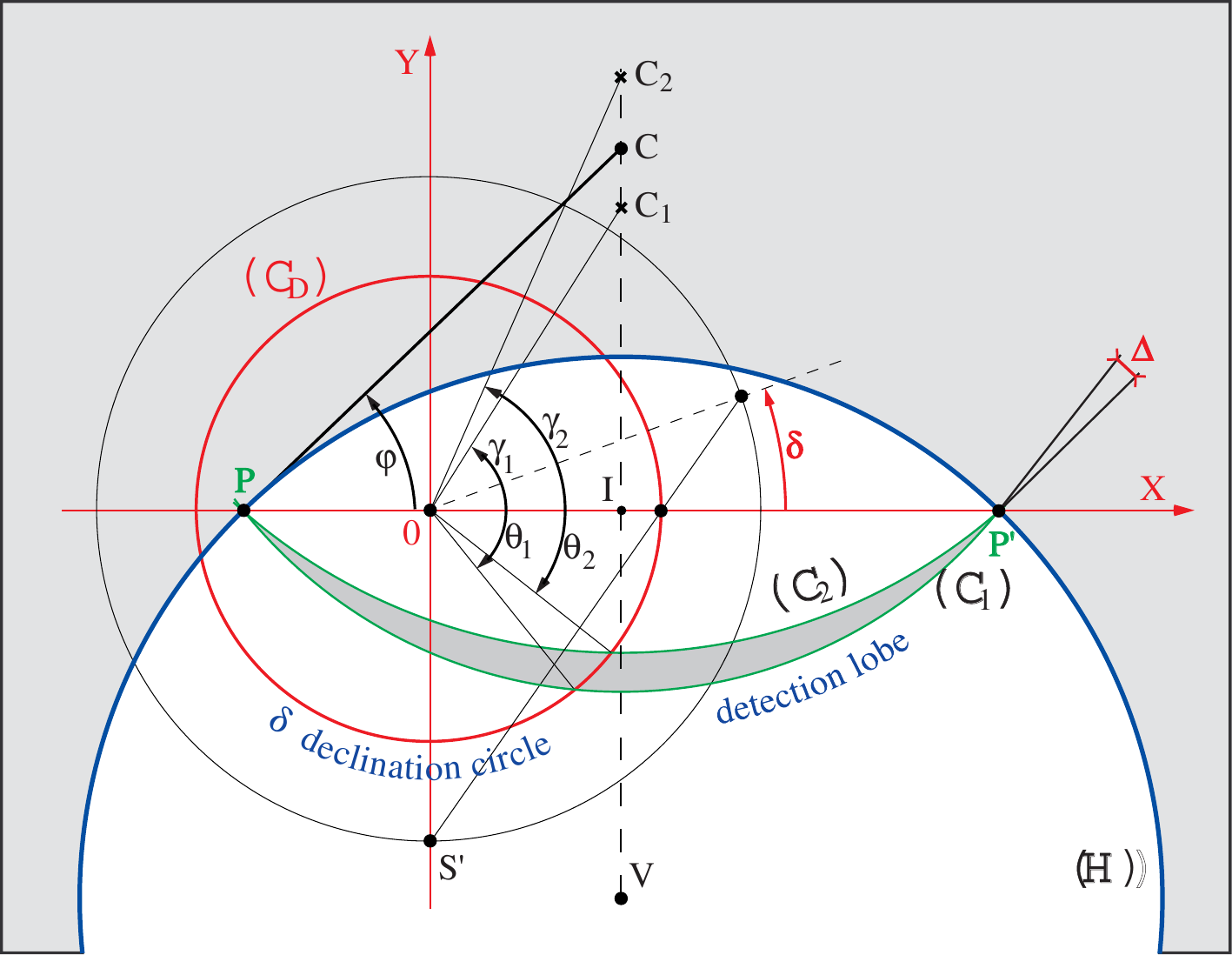}
}
\caption[]{\it The projections of the detection lobe (in grey), the horizon
(${\cal H}$, blue circle)
and a declination circle (${\cal C}_D$ in red) in the rotated
frame (see text). The radius of the
declination circle is constructed from $S'$ and $\delta$
similarly to the construction of $OM$ from $S$ and $\lambda$
in
Fig. 5.
}
\label{newref}
\end{figure}

The final calculations are made in the rotated frame (XoY) as shown in
Fig. \ref{newref}.
The parameters we will use are $OI$, $\Delta$ and $\phi$ that is given by
\begin{equation}
\tan \phi =IC/IP=\frac{\tan \lambda }{\sin A}
\end{equation}
(from (\ref{IC}) and (\ref{IP})).
The angle $\Delta$ between the lobe's side circles
is invariant under the projection, and is shown in Fig. \ref{newref}.
To find the fraction of a sideral day spent by an object at declination $\delta$
within the lobe acceptance, one needs to find the intersections of the
declination circle (${\cal C}_D$) with the images of the large circles
(${\cal C}_1$) of center $C_1\,(X_1,Y_1)$ and (${\cal C}_2$) of center
$C_2\,(X_2,Y_2)$. The equations of (${\cal C}_D$) and (${\cal C}_1$) are:
\begin{eqnarray}
X^2+Y^2&=&\tan^2(\pi/4-\delta/2) \label{D}\\
(X-X_1)^2+(Y-Y_1)^2&=&PC_1^2 \label{C1}
\end{eqnarray}
\begin{eqnarray}
(\ref{C1}) &=>& X^2+Y^2-2(XX_1+YY_1)+X_1^2+Y_1^2=PC_1^2 \\
&=>& \tan^2(\pi/4-\delta/2)-2(XX_1+YY_1)+OI^2+IC_1^2=PC_1^2=PI^2+IC_1^2 \\
&=>& \tan^2(\pi/4-\delta/2)-2(XX_1+YY_1)=PI^2-OI^2=1\ (from\ (\ref{IP}) \label{tan})
\end{eqnarray}
using polar coordinates
\begin{eqnarray}
X=R.\cos \theta_1 & & X_1=OC_1.\cos \gamma_1 \\
Y=R.\sin \theta_1 & & Y_1=OC_1.\sin \gamma_1
\end{eqnarray}
\begin{eqnarray}
(\ref{D})&=>& R=\tan(\pi/4-\delta/2)\ ({\rm positive\ because}\ -\pi/2<\delta <\pi/2)\ {\rm and} \\
(\ref{tan})&=>&\tan^2(\pi/4-\delta/2)-2\tan (\pi/4-\delta/2)\times OC_1.(\cos\theta_1 \cos\gamma_1 +\sin \theta_1 \sin \gamma_1)=1\\
&=>& \tan^2(\pi/4-\delta/2)-2\tan (\pi/4-\delta/2)\times OC_1.\cos(\theta_1 -\gamma_1)=1 \\
&=>& \cos(\theta_1 -\gamma_1)=\frac{\tan^2(\pi/4-\delta/2)-1}{2\tan (\pi/4-\delta/2)\times OC_1}=\frac{-\tan \delta}{OC_1}
\end{eqnarray}
using the formula of the half-angle tangent.\\
$\gamma_1$ is given by:
\begin{equation}
\tan \gamma_1 =IC_1/OI=\frac{IP\tan (\phi-\Delta/2)}{OI}=\frac{\tan (\phi-\Delta/2)}{\cos\lambda \cos A },
\end{equation}
using (\ref{OI}) and (\ref{IP}).\\
$OC_1$ is given by (using also (\ref{OI})):
\begin{equation}
OC_1^2=OI^2+IC_1^2=OI^2(1+\tan^2 \gamma_1)
=\frac{\cos^2 \lambda }{\sin^2 \lambda+\tan^2 A}\left[1+\frac{\tan^2(\phi-\Delta/2)}{\cos^2 \lambda \cos^2 A }\right]
\end{equation}
which can be written:
\begin{equation}
OC_1^2=\frac{\cos^2 \lambda \cos^2 A +\tan^2(\phi-\Delta/2)}{1-\cos^2 \lambda \cos^2 A }.
\end{equation}
It follows
\begin{equation}
cos(\theta_1 -\gamma_1)=-\tan \delta \sqrt{\frac{1-\cos^2 \lambda \cos^2 A }{\cos^2 \lambda \cos^2 A +\tan^2(\phi-\Delta/2)}}.
\end{equation}
\fbox{
\begin{minipage}{1.0\textwidth}
{\bf The expression for the searched angles is given by (0, 1 or 2 solutions):}\\
\begin{equation}
\theta_1=\arctan\left[\frac{\tan(\phi-\Delta/2)}{\cos\lambda \cos A }\right]
\pm \arccos\left[-\tan \delta\sqrt{\frac{1-\cos^2\lambda\cos^2 A}{\cos^2 \lambda \cos^2 A +\tan^2(\phi-\Delta/2)}}\right]
\label{theta}
\end{equation}
where \\
\begin{equation}
\tan \phi =\frac{\tan \lambda}{\sin A}\ =>\ \tan(\phi-\Delta/2)=\frac{\tan\phi - \tan(\Delta/2)}{1 + \tan\phi \tan(\Delta/2)}=\frac{\tan\lambda-\sin A \tan(\Delta/2)}{\sin A + \tan\lambda\tan(\Delta/2)}
\end{equation}
{\it Choice of determinations:}\\
- The fact that $\lambda>0$ and $0<A<\pi/2$ implies that the
determination of the $\arctan$ (for angle $\gamma_1$) is between $-\pi/2$ and $+\pi/2$.\\
- As the two solutions correspond to the two determinations for the $\arccos$,
the choice of the first determination can be made between $0$ and $\pi$.
\end{minipage}
}
\\
\\
{\bf Exchanging $-\Delta$ into $+\Delta$ and $\gamma_1$ into $\gamma_2$
gives the corresponding result for $\theta_2$.}

\subsection{Conditions of observability}
An object with declination $\delta$ is observable if \\
i) there is a solution for $\theta_1$ or $\theta_2$, and if \\
ii) this solution corresponds to a configuration
above horizon, {\it i.e.} if the associated point on the sphere of
Fig. \ref{vision3D} belongs to the half-sphere of pole $M_0$.
The stereographic projection of the limit of this half-sphere is the
horizon circle (${\cal H}$).
The intersections of (${\cal C}_D$) with (${\cal C}_1$) or (${\cal C}_2$)
on the projection
correspond to the visibility limits if they are inside (${\cal H}$), that
contains the hatched lobe defined by $P$ $P'$ and $M$.\\
- The first condition (existence of at least one solution) can be expressed by:
\begin{equation}
\vert\tan \delta\vert<\sqrt{\frac{\cos^2 \lambda \cos^2 A +\tan^2(\phi+\Delta/2)}{1-\cos^2\lambda\cos^2 A}}
\end{equation}
or equivalently
\begin{equation}
(1-\cos^2\lambda\cos^2 A)\tan^2\delta <\cos^2\lambda \cos^2 A +\tan^2(\phi+\Delta/2)
\end{equation}
- The second condition (visibility)
is satisfied if the intersection is
within the disk centered on $V$ with radius $VP=1/\sin\lambda$,
corresponding to the inequality relation:
\begin{equation}
(X-OI)^2+(Y+IV)^2<1/\sin^2\lambda ,
\end{equation}
with $IV=OI\tan\alpha=OI\tan A/\sin\lambda$ (from Fig.\ref{proj2D} and (\ref{tgalpha})). \\
In polar coordinates $(R,\theta)$, this condition, applied to the intersection points,
becomes
\begin{equation}
X^2+Y^2+2(Y.IV-X.OI)+OV^2<1/\sin^2\lambda\ <=>
\end{equation}
\begin{equation}
\tan^2(\pi/4-\delta/2)+2\tan(\pi/4-\delta/2)
\frac{\cos\lambda}{\sqrt{\sin^2 \lambda+\tan^2 A}}
(\frac{\tan A}{\sin\lambda}\sin\theta-\cos\theta)
+\cot^2\lambda<1/\sin^2\lambda
\end{equation}
using (\ref{OI}), (\ref{OV}) and $R=\tan(\pi/4-\delta/2)$.

After simplification, one gets:
\begin{equation}
(\frac{\tan A}{\tan\lambda}\sin\theta-\cos\lambda\cos\theta)
\frac{1}{\sqrt{\sin^2 \lambda+\tan^2 A}}
<\frac{1-\tan^2(\pi/4-\delta/2)}{2\tan (\pi/4-\delta/2)}= \tan\delta
\end{equation}
using again the half-angle tangent formula.
As $\lambda>0$, one obtains finally the condition:
\begin{equation}
\tan A\sin\theta-\sin\lambda\cos\theta < \tan\delta\tan\lambda \sqrt{\sin^2\lambda+\tan^2 A}.
\label{Condition}
\end{equation}
\subsection{Exposure time calculation}
After establishing the list of lobe-crossings that are visible (above
horizon), one has to distinguish different relative configurations
of the declination circle with respect to the lobe (the illustrations of
the next section may help the reader at this stage) :
\begin{itemize}
\item
{\bf No lobe-crossing (0 solution):}
The declination circle is completely inside or outside the
lobe (shaded area).
Assuming $\lambda>0$,
the daily exposure is 24 sideral hours if $\delta>0$ and if the North
pole (projection $O$) is within the lobe.
The pole is within the lobe if $C$ is not between $C_1$
and $C_2$ (see Fig. \ref{newref}), condition expressed by:
\begin{equation}
(\tan \phi-\tan(\phi-\Delta/2))\times (\tan \phi-\tan(\phi+\Delta/2))>0.
\end{equation}
Otherwise, the exposure time is zero.
\item
{\bf Lobe-crossings happens and the pole is NOT in the lobe:}
The exposure time at a given latitude is
obtained by ordering the list of $0<\theta_1<2\pi$ and $0<\theta_2<2\pi$ values
that satisfy the visibility condition (\ref{Condition}) by increasing
order ($\theta(i)$, i=1 to 4 at maximum) from zero, and account for the value
$(\theta(i+1)-\theta(i))/2\pi \times 1\, sideral \, day$ per lobe-crossing.
\item
{\bf Lobe-crossings happens and the pole is in the lobe:}
In this case the $\theta=0$ point of the declination circle is within
the detection lobe. The list of $\theta_1$ and $\theta_2$ that satisfy
the visibility conditions has to start with the largest value
(between $0$ and $2\pi$), followed by the others by increasing order
from zero. Then the exposure time is obtained by the sum of values
$((\theta(i+1)-\theta(i))/2\pi \times 1\, sideral \, day$ starting from $i=1$.
\end{itemize}

\section{Some particular cases}
\begin{itemize}
\item
$A=0$, antenna oriented North-South (Fig. \ref{casns}a).\\
\begin{figure}[h]
\centering
\vbox{
\includegraphics[width=8cm]{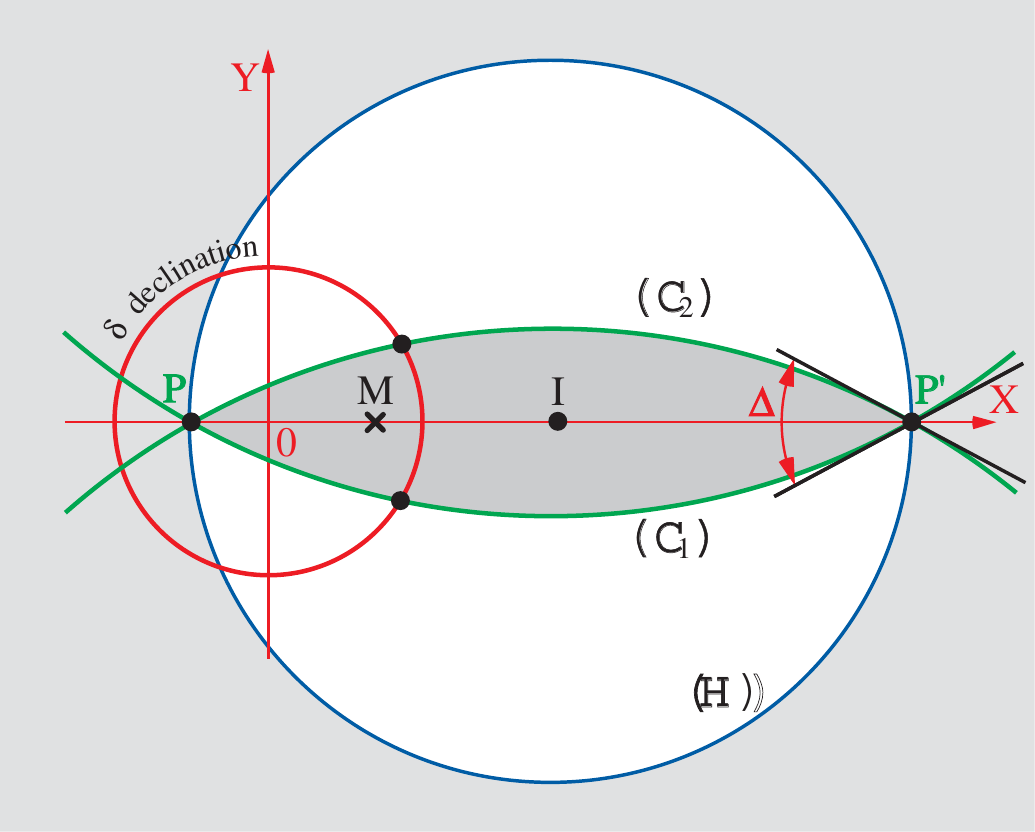}\includegraphics[width=7cm]{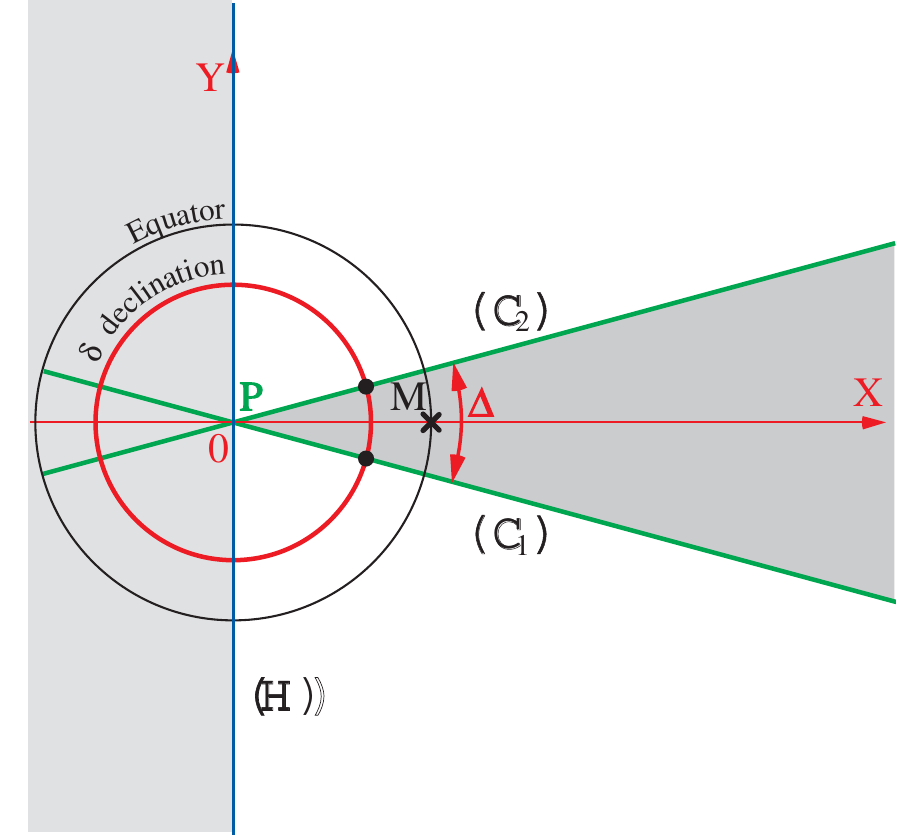}
}
\caption[]{\it
(a) The projected lobe when the antenna is oriented North-South ($A=0^\circ$).\\
(b) The particular case of the equatorial location ($\lambda=0^\circ$ and $A=0^\circ$).
}
\label{casns}
\end{figure}

$\phi=\pi/2$ and (\ref{theta}) simplifies into
\begin{equation}
\theta_1=\arctan\left[\frac{\cot(\Delta/2)}{\cos\lambda}\right]
\pm \arccos\left[\frac{-\tan \delta \sin\lambda}{\sqrt{\cos^2 \lambda +\cot^2(\Delta/2)}}\right]
\end{equation}
- If $\delta<\lambda-\pi/2$, the object is not visible. \\
- If $\lambda-\pi/2<\delta<\pi/2-\lambda$,
then the object enters the visibility lobe once per day during the exposure time
\begin{equation}
t_{exp}=\frac{1\ day}{\pi}
\left[-\arctan\left[\frac{\cot(\Delta/2)}{\cos\lambda}\right]+
\arccos\left[\frac{-\tan\delta \sin\lambda}{\sqrt{\cos^2 \lambda +\cot^2(\Delta/2)}}\right]\right]
\end{equation}
using the positive determinations for the $arctan$ and the $arccos$ in this expression.\\
- If $\delta>\pi/2-\lambda$ and
$\tan\delta<\frac{\sqrt{\cos^2\lambda+\cot^2(\Delta/2)}}{\sin\lambda}$,
the object is circumpolar and enters the visibility lobe twice per day
during the total exposure time given by:
\begin{eqnarray}
t_{exp}=\frac{1\ day}{\pi}
\left[2\arccos\left[\frac{-\tan\delta \sin\lambda}{\sqrt{\cos^2 \lambda +\cot^2(\Delta/2)}}\right]-\pi \right]\nonumber \\
=\frac{1\ day}{\pi}
\left[\pi-2\arccos\left[\frac{\tan\delta \sin\lambda}{\sqrt{\cos^2 \lambda +\cot^2(\Delta/2)}}\right]\right].
\end{eqnarray}
- If $\tan\delta>\frac{\sqrt{\cos^2\lambda+\cot^2(\Delta/2)}}{\sin\lambda}$
(which is close to the condition $\delta>\pi/2-\Delta/2$ if $\Delta$ is small),
the object is near the pole and is always in the visibility lobe.
\item
If $A=0$ and $\lambda=0$ (antenna on the equator),
then the lobe is defined by two half-lines (Fig. \ref{casns}b), and the
full sky is visible with uniform daily exposures
$t_{exp}=\Delta/2\pi \times 1\, sideral\, day$.
\item
$A=\pi/2$, antenna oriented East-West (Fig. \ref{caseo}a).\\
\begin{figure}[h]
\centering
\vbox{
\includegraphics[width=7cm]{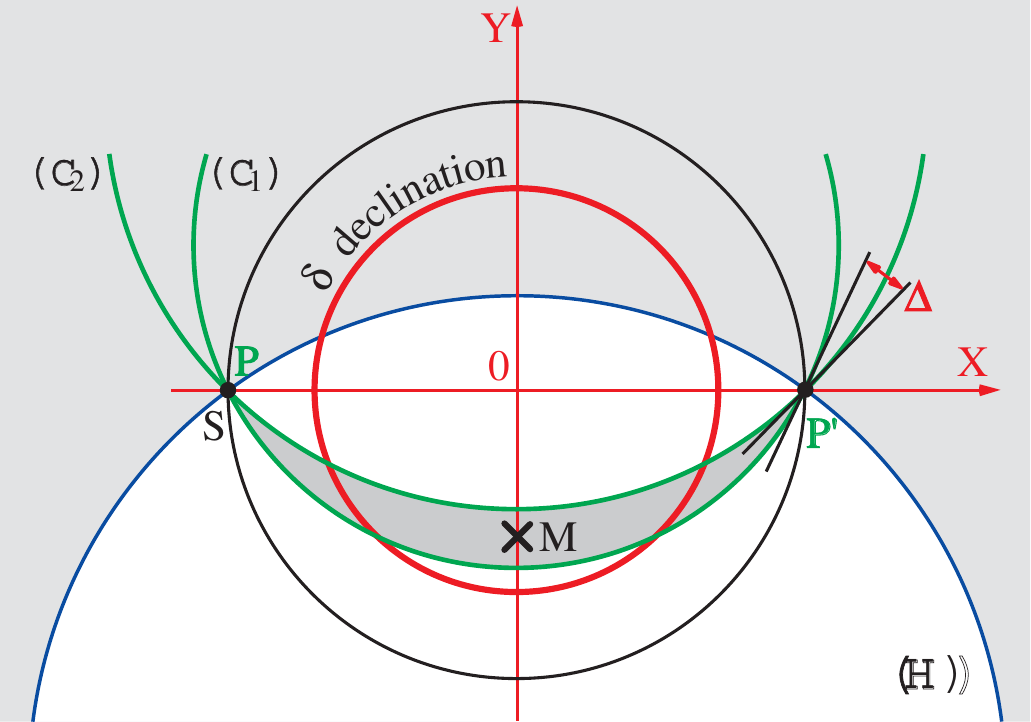}\includegraphics[width=7cm]{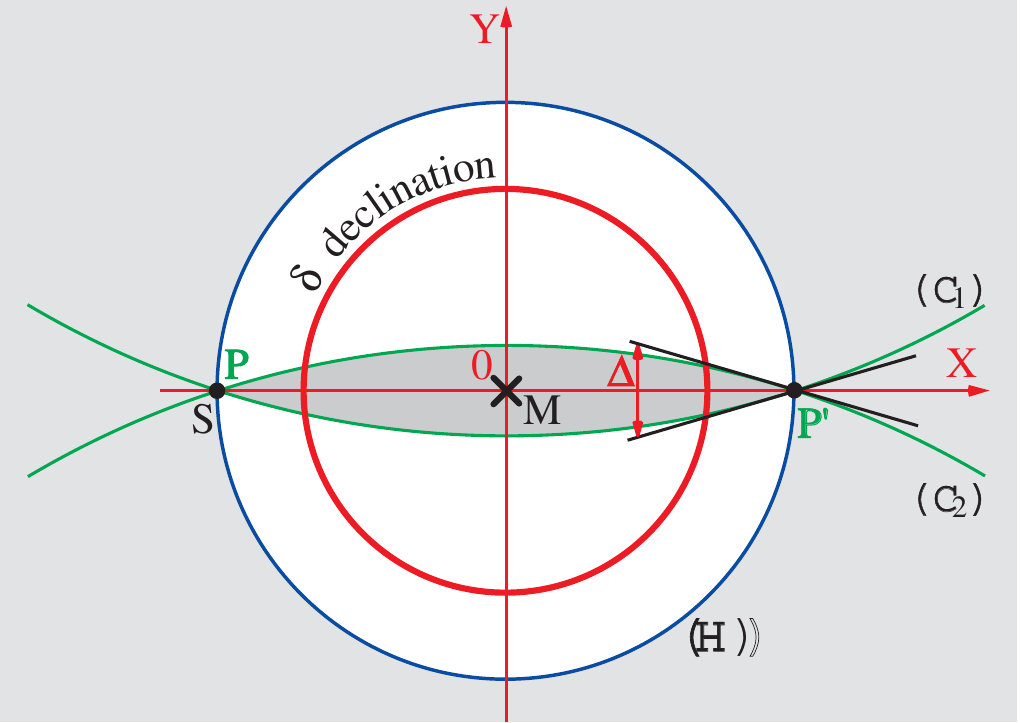}
}
\caption[]{\it
(a) The projected lobe when the antenna is oriented East-West ($A=\pi/2$).\\
(b) The particular case of the polar location ($\lambda=\pi/2$ and $A=\pi/2$).
}
\label{caseo}
\end{figure}

$\phi=\lambda$ and (\ref{theta}) simplifies into
\begin{equation}
\theta_1=\pi/2
\pm \arccos\left[\frac{-\tan \delta}{\tan(\lambda-\Delta/2)}\right]
\end{equation}
From Fig.\ref{caseo}a we find that the visibility conditions are simply
$$min(0,\lambda - \Delta/2) < \delta < \lambda + \Delta/2.$$
\item
$\lambda=\pi/2$ (antenna at the north pole).\\
The lobe can be seen on Fig. \ref{caseo}b.
$\phi=\pi/2$ and (\ref{theta}) simplifies into
\begin{equation}
\theta_1=\pi/2
\pm \arccos\left[-\tan \delta .\tan(\Delta/2)\right].
\end{equation}
If $\delta>\pi/2-\Delta/2$ then the object is always in the lobe;
if $\delta<\pi/2-\Delta/2$ then the daily exposure is given by
$t_{exp}=(1-\frac{2}{\pi}\arccos\left[\tan \delta
\tan(\Delta/2)\right])\times (1\, sideral\, day)$.
\end{itemize}

\section{Study of configurations}
In this section, we study the impact of several orientations of a
static cylindrical reflector on the field coverage and on the daily
exposure time. BAO low-z studies should benefit from the widest
coverage, favoured by the North-South ($A=0$) orientation of the
cylinder axis; but high-z studies would need long exposures
of low synchrotron background temperature fields, to
allow deeper observations, a sensitivity that could be easier 
to reach with a different axis orientation.

\begin{figure}[h]
\centering
\vbox{
\includegraphics[width=14cm]{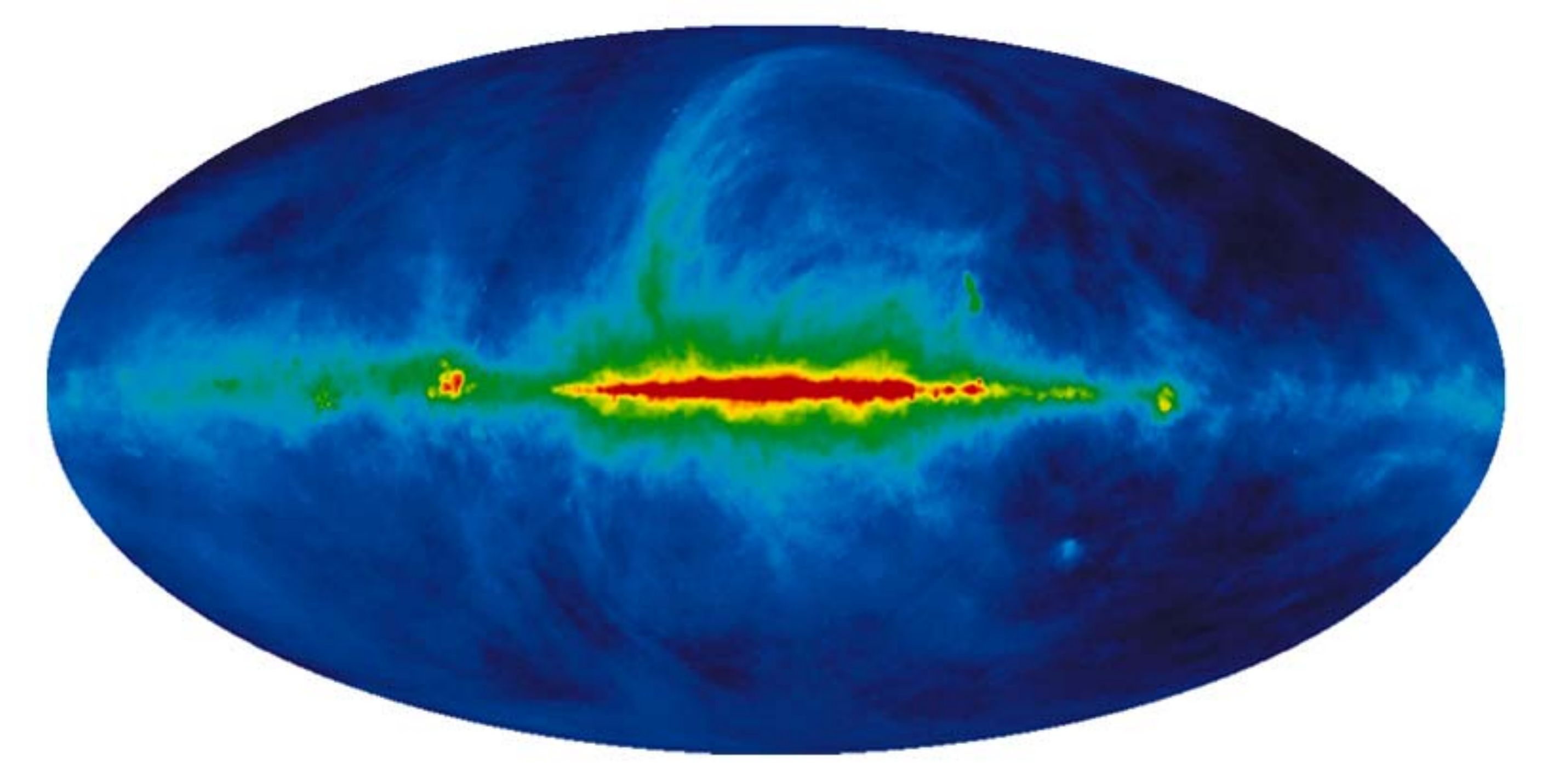}
}
\caption[]{\it
The Haslam map of the synchrotron galactic emission at 408 MHz
(galactic coordinates).
}
\label{synchro}
\end{figure}

Fig. \ref{synchro} shows the map of foreground galactic synchrotron emission\cite{4}
at $\sim 74 cm$ \footnote{http://lambda.gsfc.nasa.gov/product/foreground}.
This foreground has to be considered together with the exposure maps
given below, in order to optimize the position and azimuth of the antenna.

\subsection{Nan\c{c}ay}
\begin{figure}[h]
\centering
\begin{minipage}{7.0cm}
\includegraphics[width=6.5cm]{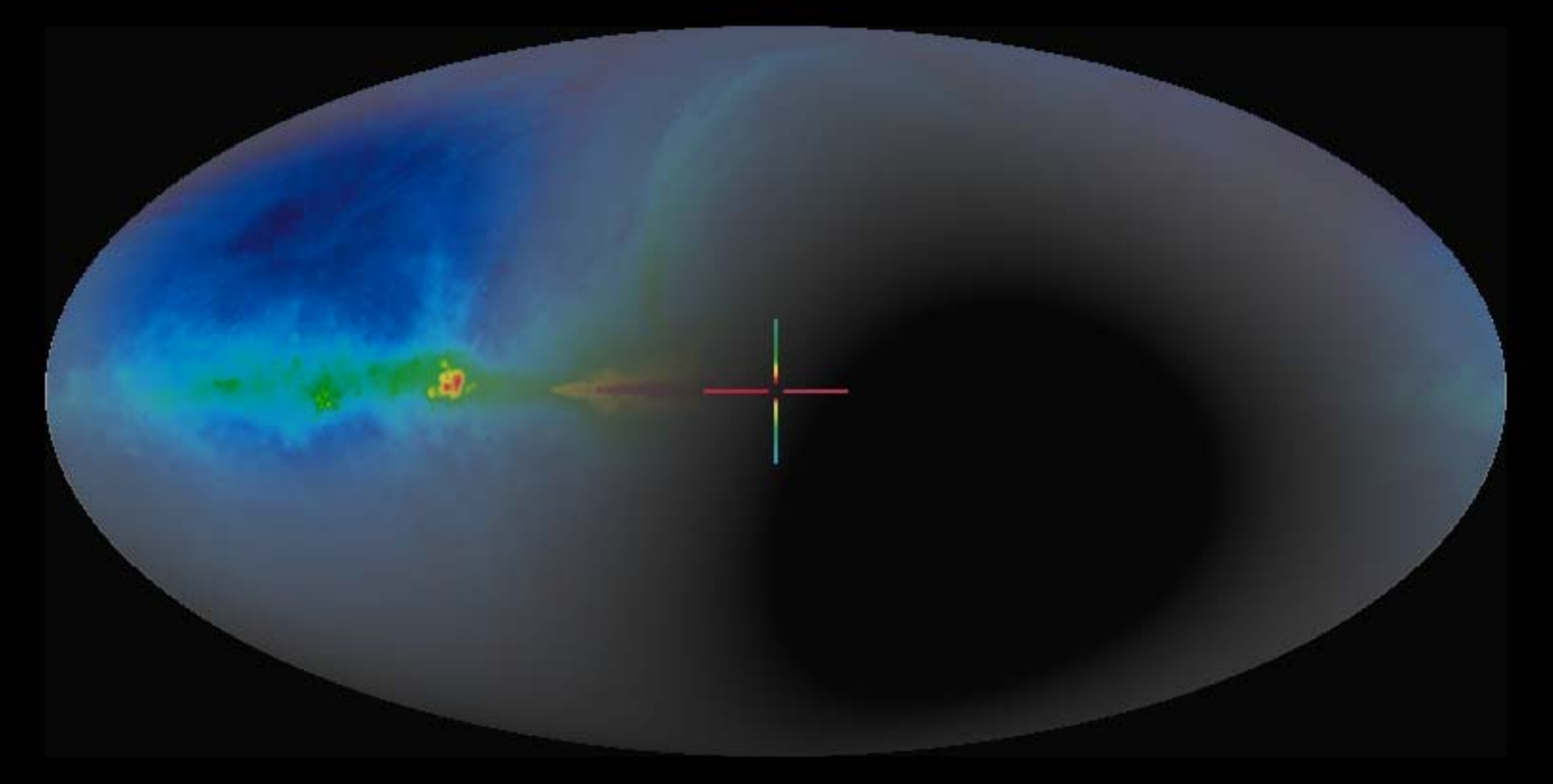}

\includegraphics[width=6.5cm]{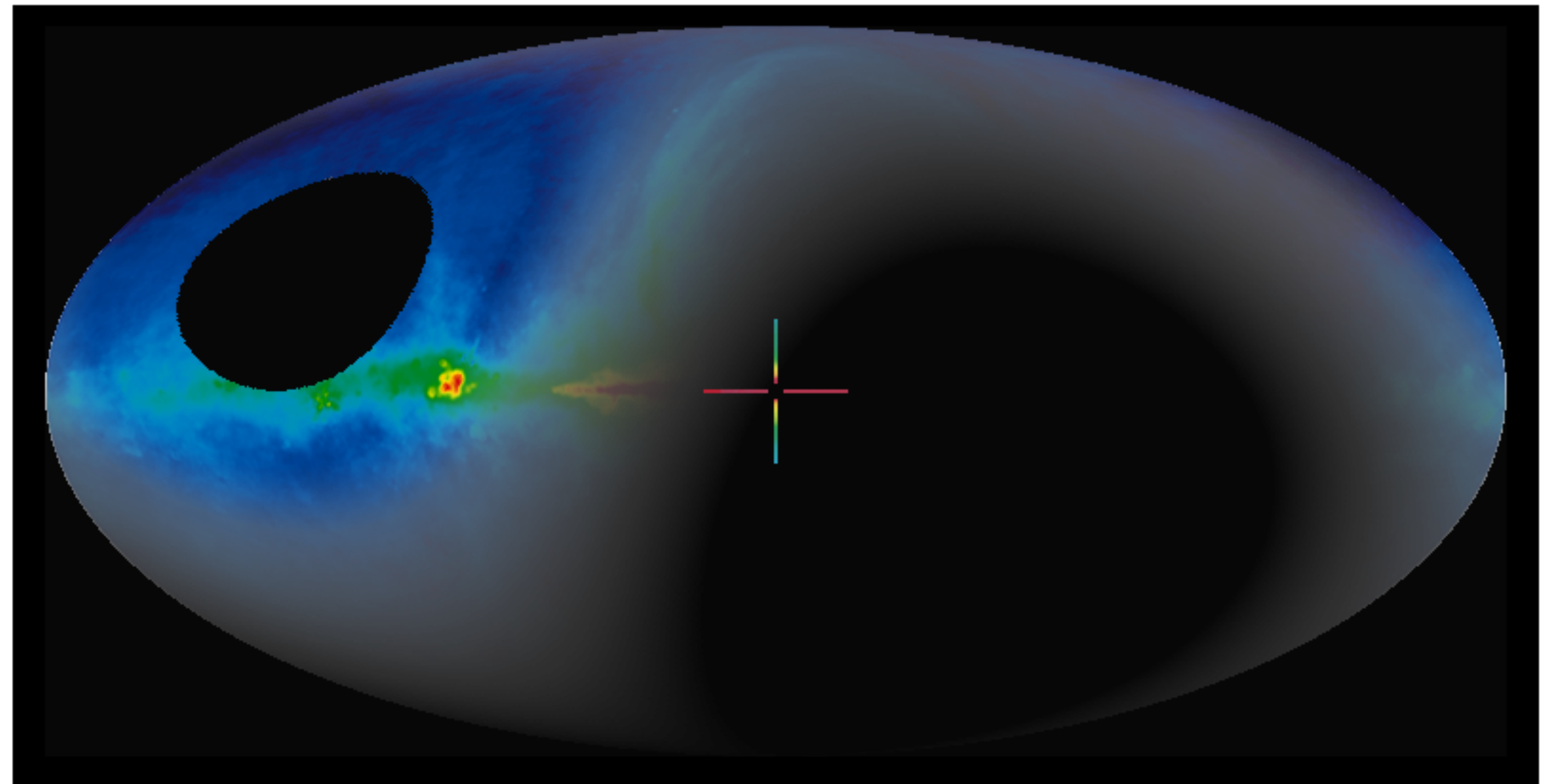}

\includegraphics[width=6.5cm]{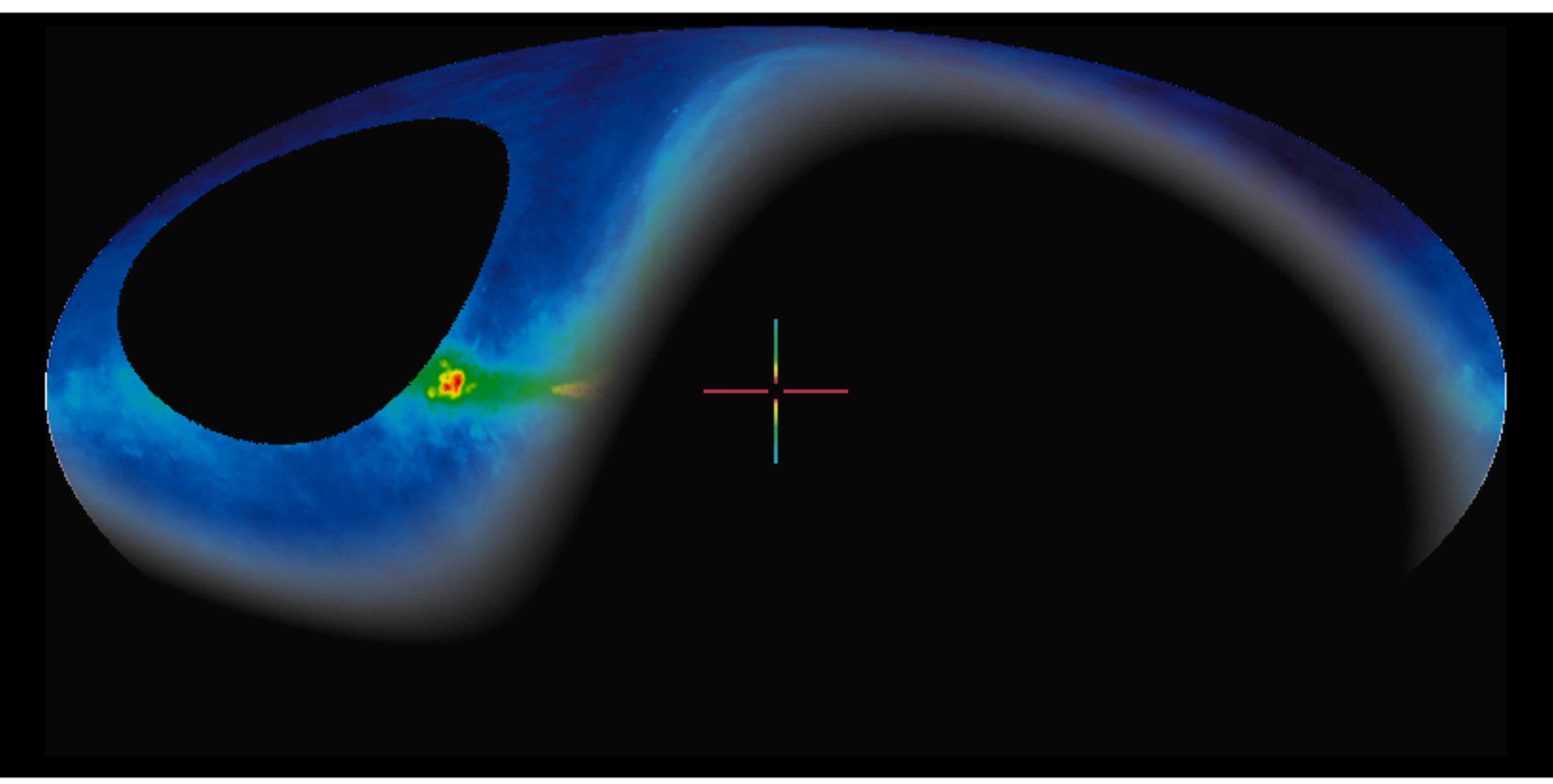}
\end{minipage}
\begin{minipage}{7.5cm}
\includegraphics[width=6.5cm,height=4.5cm]{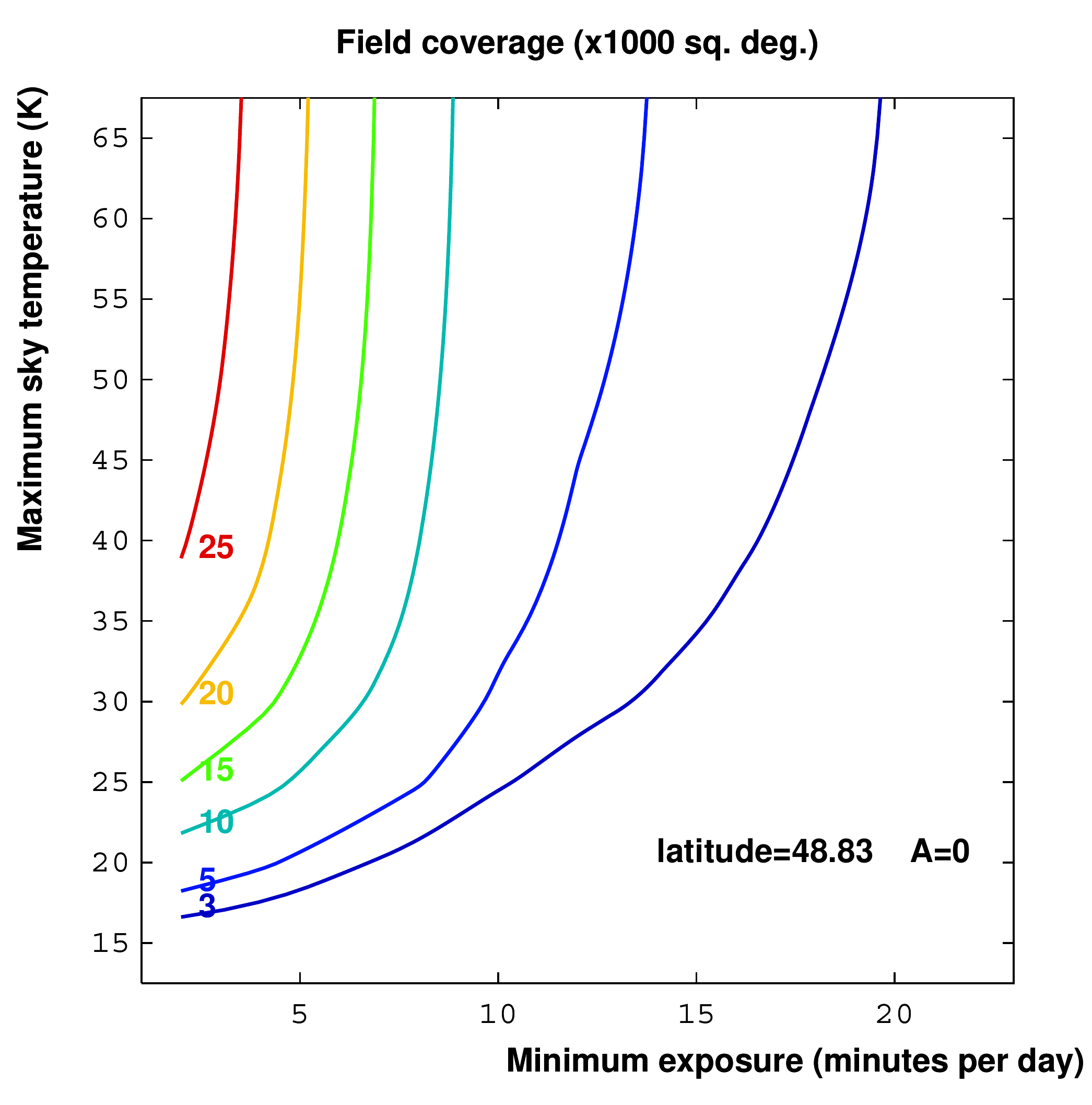}

\includegraphics[width=6.5cm,height=4.5cm]{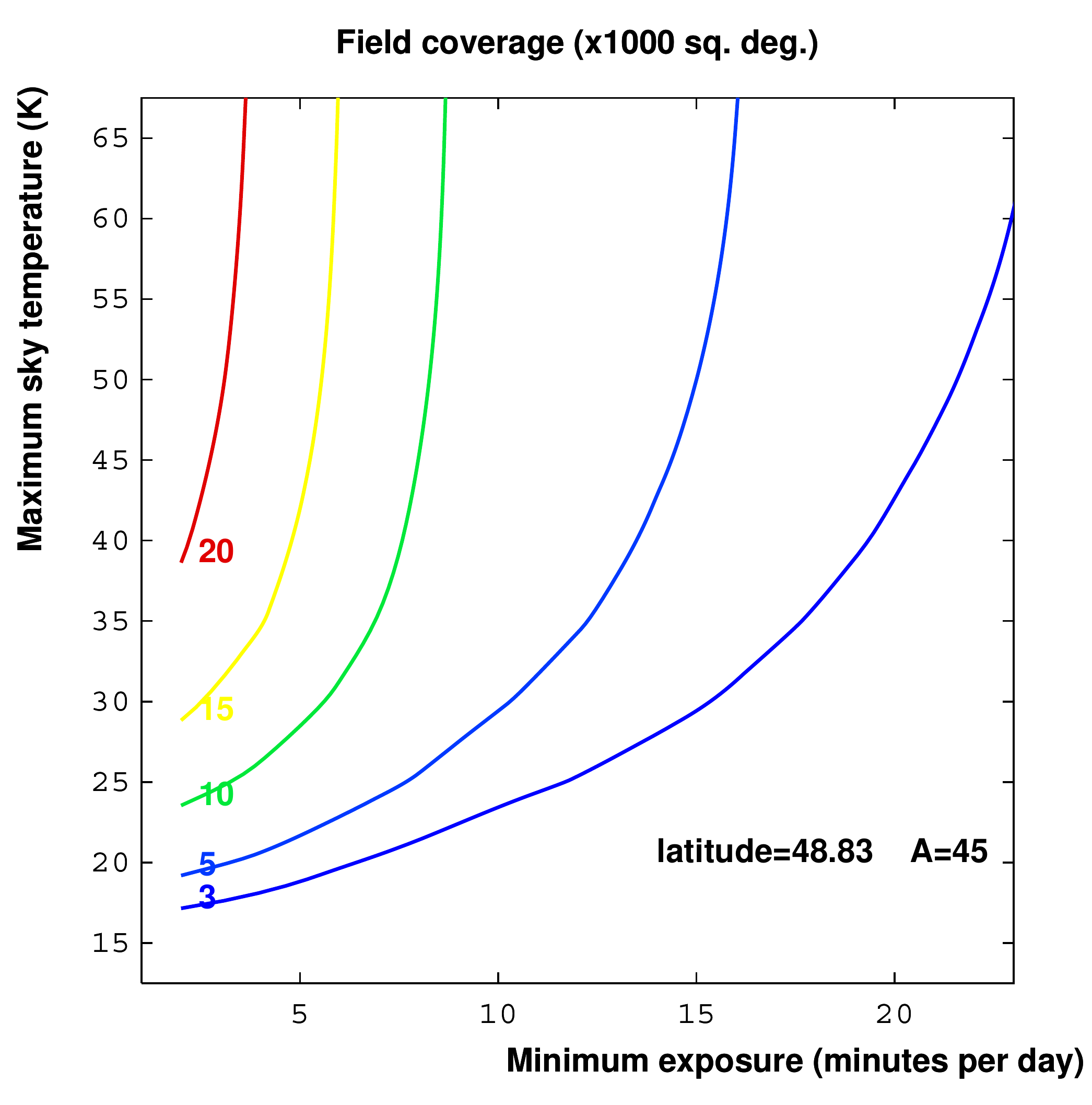}

\includegraphics[width=6.5cm,height=4.5cm]{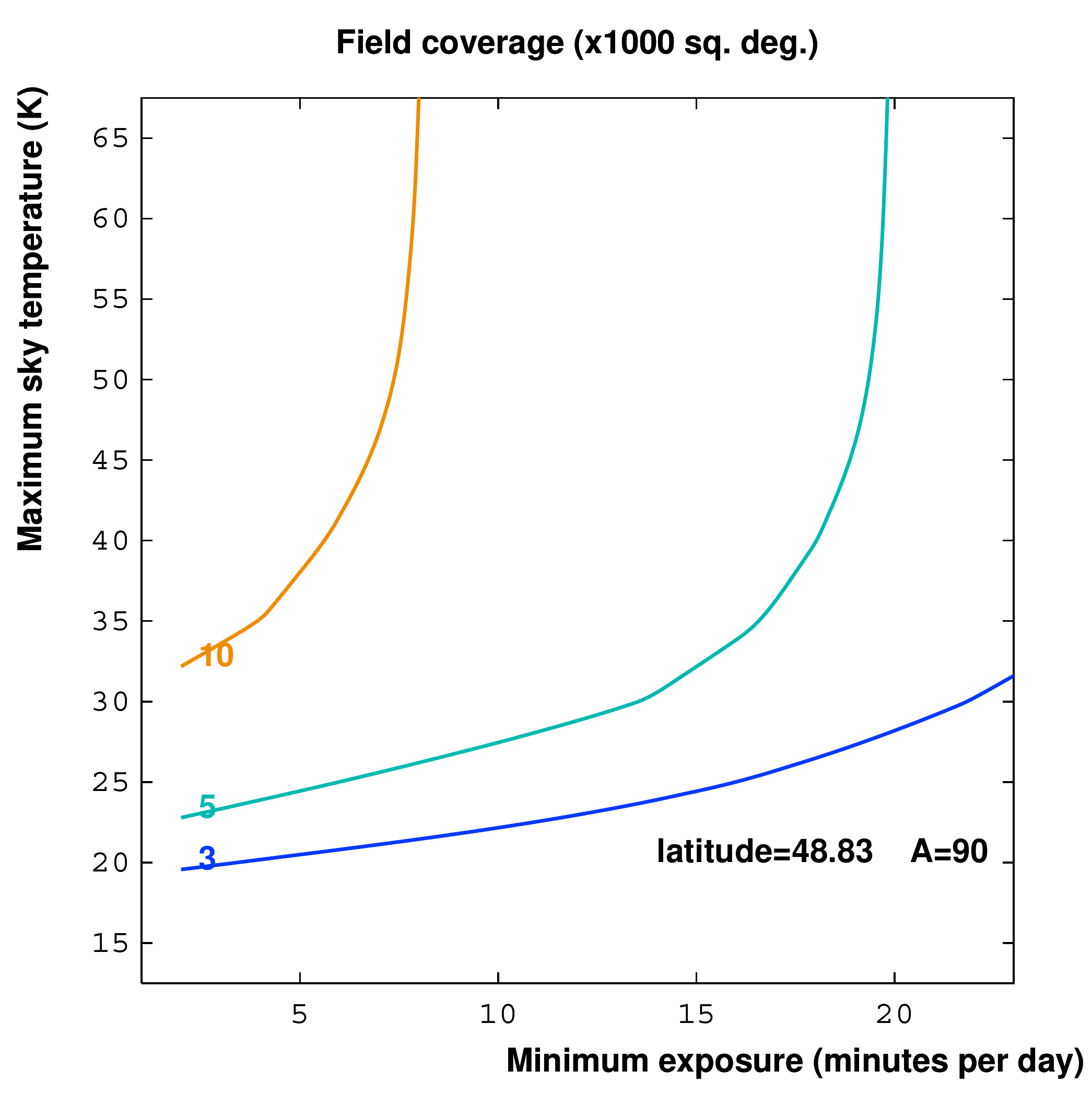}
\end{minipage}
\caption[]{
\it
Antenna located at Nan\c{c}ay latitude (France).

LEFT:
Sky visibility (luminosity proportional to the Exposure time)
as a function of galactic coordinates.

From top to bottom:
antenna azimuth $A=0^\circ$ (North-South), $A=45^\circ$ and
$A=90^\circ$ (East-West).

RIGHT:
field covered by the antenna as a function of the minimum daily exposure
and the maximum synchrotron sky temperature.
For example, for $A=45^\circ$ (center), the point of coordinates
$(15min.,30K)$ on the 3000 square
degree curve means that 3000 square degrees of field with a sky
background temperature below $30K$ transit during more than $15 min.$
per sideral day in the instrumental lobe.
These series of curves allows one to study the compromise between the field
of view, the background level, and the exposure time.
}
\label{nancay}
\end{figure}

Fig. \ref{nancay} (left) gives the exposure time for an antenna
with a $\Delta=2^\circ$ lobe, located
at Nan\c{c}ay (France) as a function of the
galactic coordinates for different orientations.
Fig. \ref{nancay} (right) gives the field covered by the antenna
with a daily exposure exceeding the abscissa-value and a sky synchrotron
temperature lower than the ordinate-value.
\begin{itemize}
\item
For $A=0^\circ$, 21500 square degree (52\%) of the sky are covered with a daily
exposure larger than 300s, and 2000 square degree (5\%) are covered with an
exposure larger than 1500s.
\item
For $A=45^\circ$, 17800 square degree (43\%) of the sky are covered with a daily
exposure larger than 300s, and 2800 square degree (7\%) are covered with an
exposure larger than 1500s.
\item
For $A=90^\circ$, 12200 square degree (30\%) of the sky are covered with a daily
exposure larger than 300s, and 3900 square degree (10\%) are covered with an
exposure larger than 1500s.
\end{itemize}
\subsection{Morocco}
\begin{figure}[h]
\centering
\begin{minipage}{7.0cm}
\includegraphics[width=6.5cm]{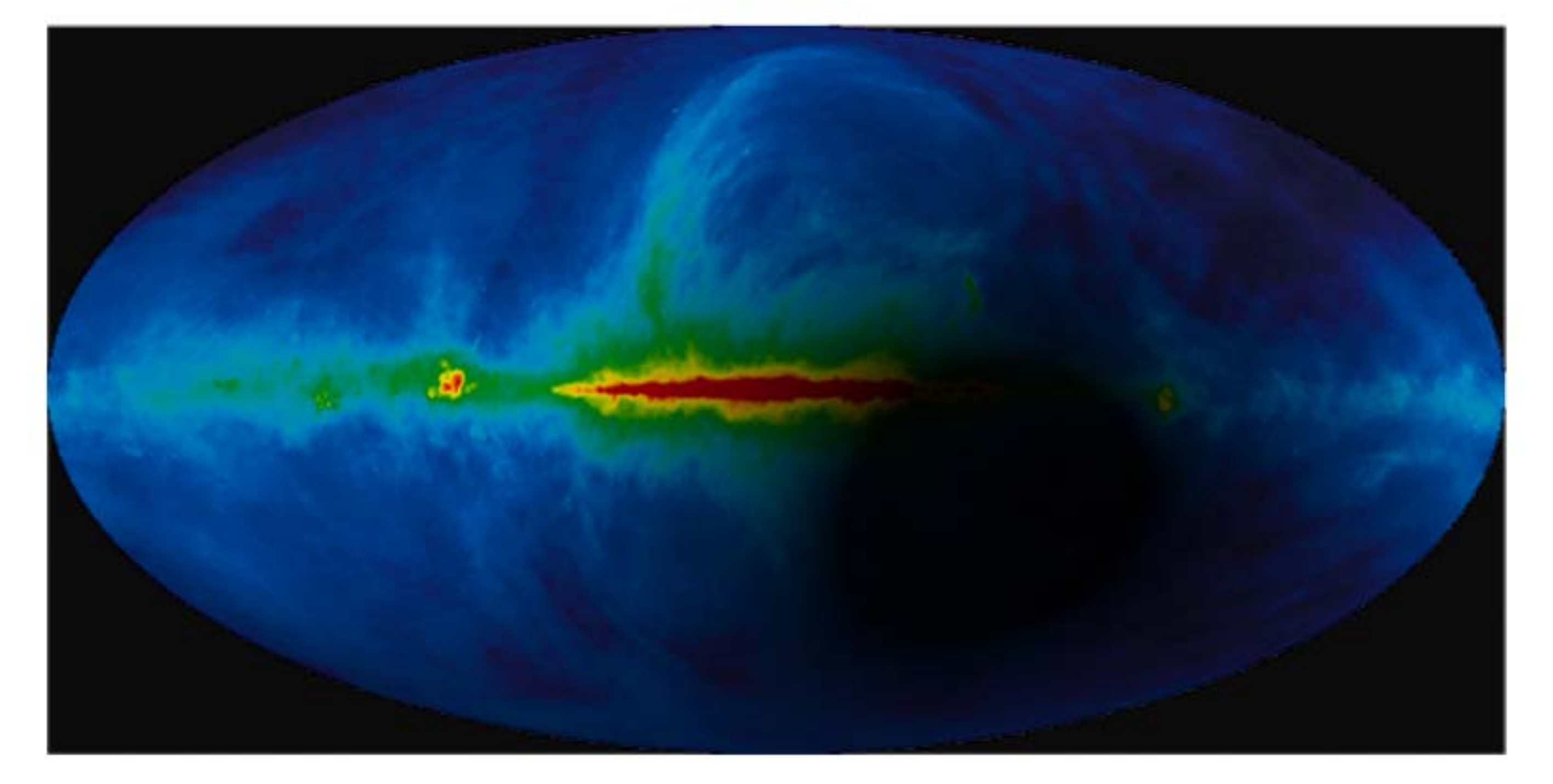}

\includegraphics[width=6.5cm]{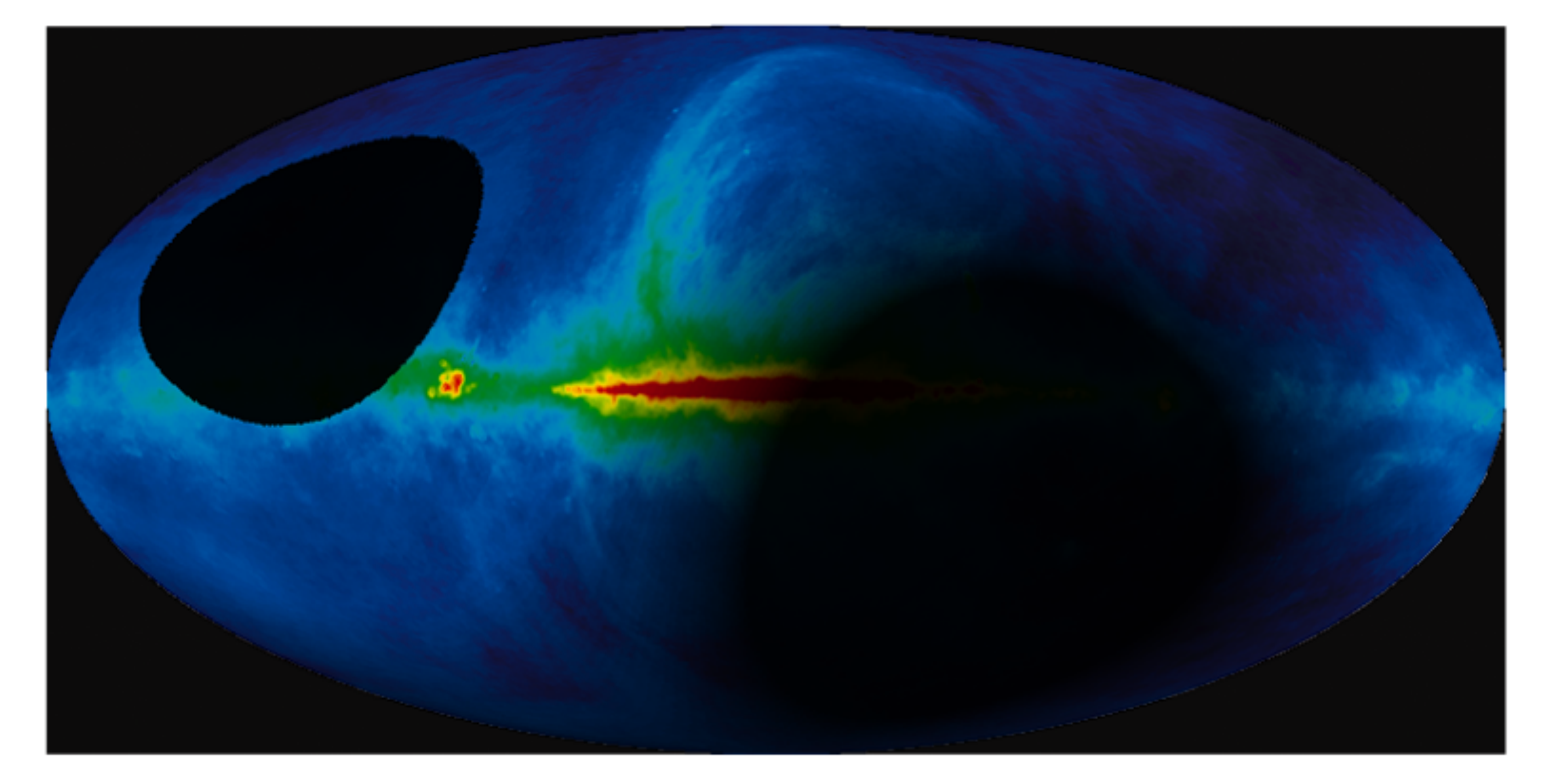}

\includegraphics[width=6.5cm]{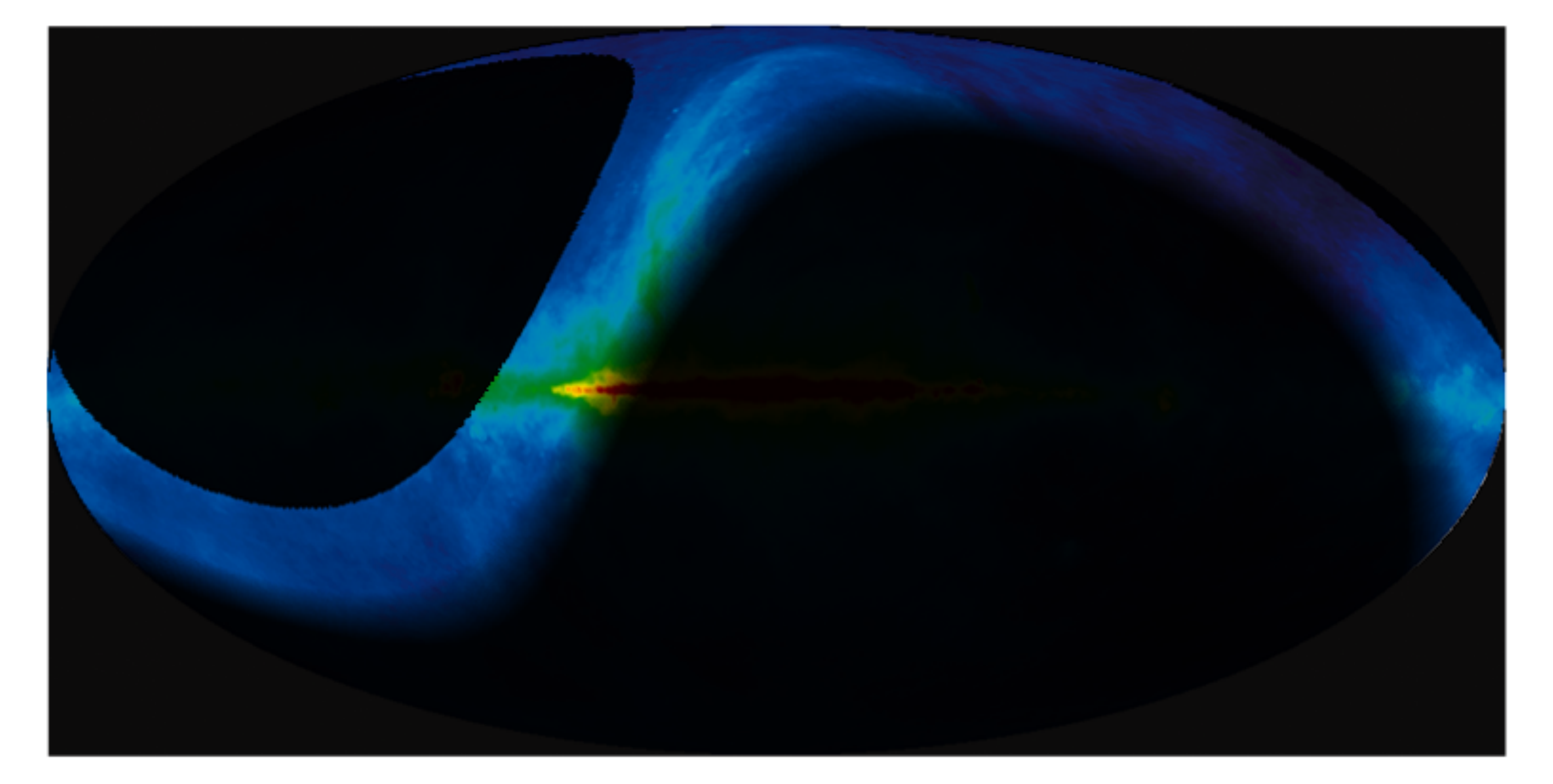}
\end{minipage}
\begin{minipage}{7.5cm}
\includegraphics[width=6.5cm,height=4.5cm]{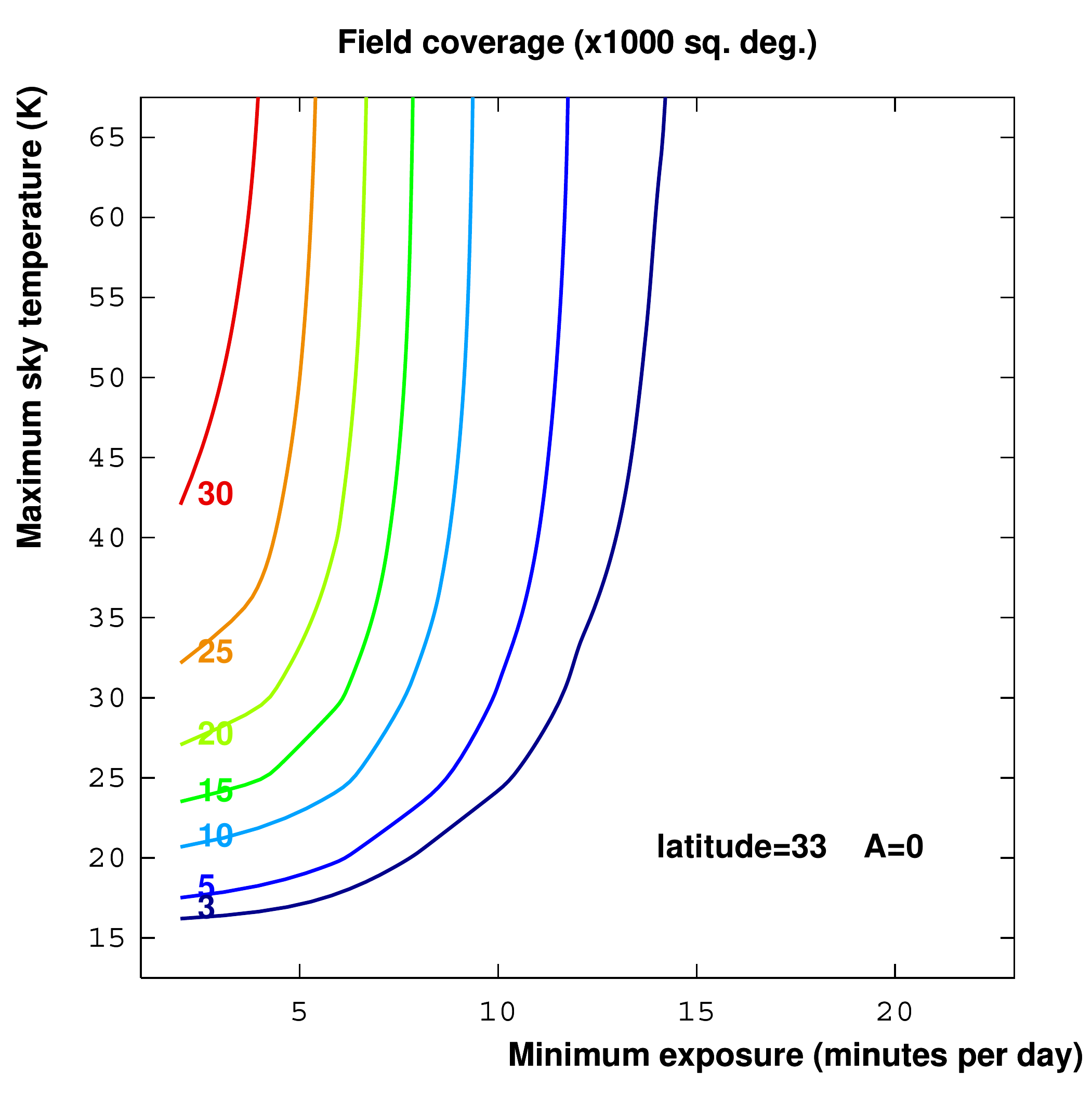}

\includegraphics[width=6.5cm,height=4.5cm]{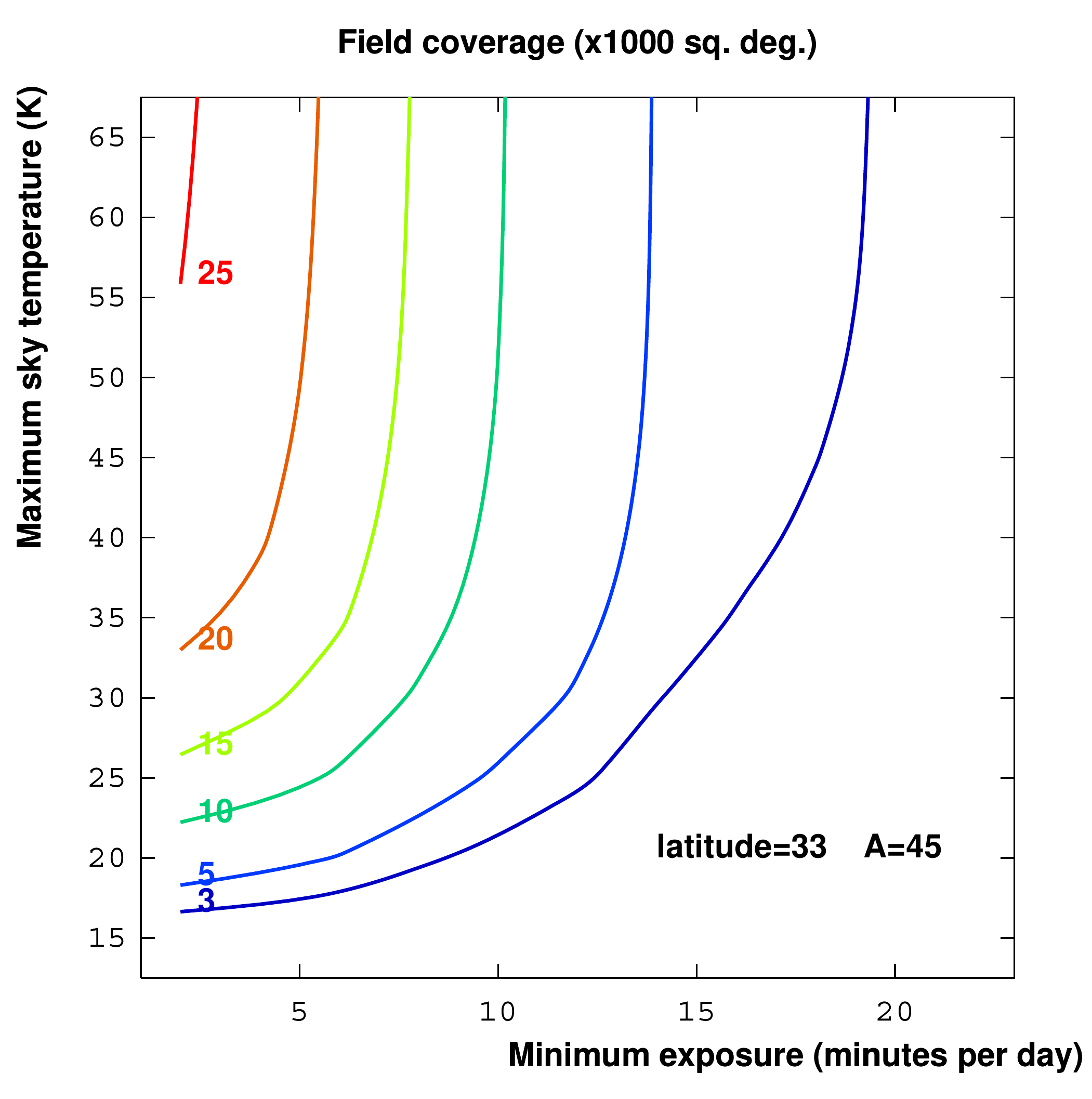}

\includegraphics[width=6.5cm,height=4.5cm]{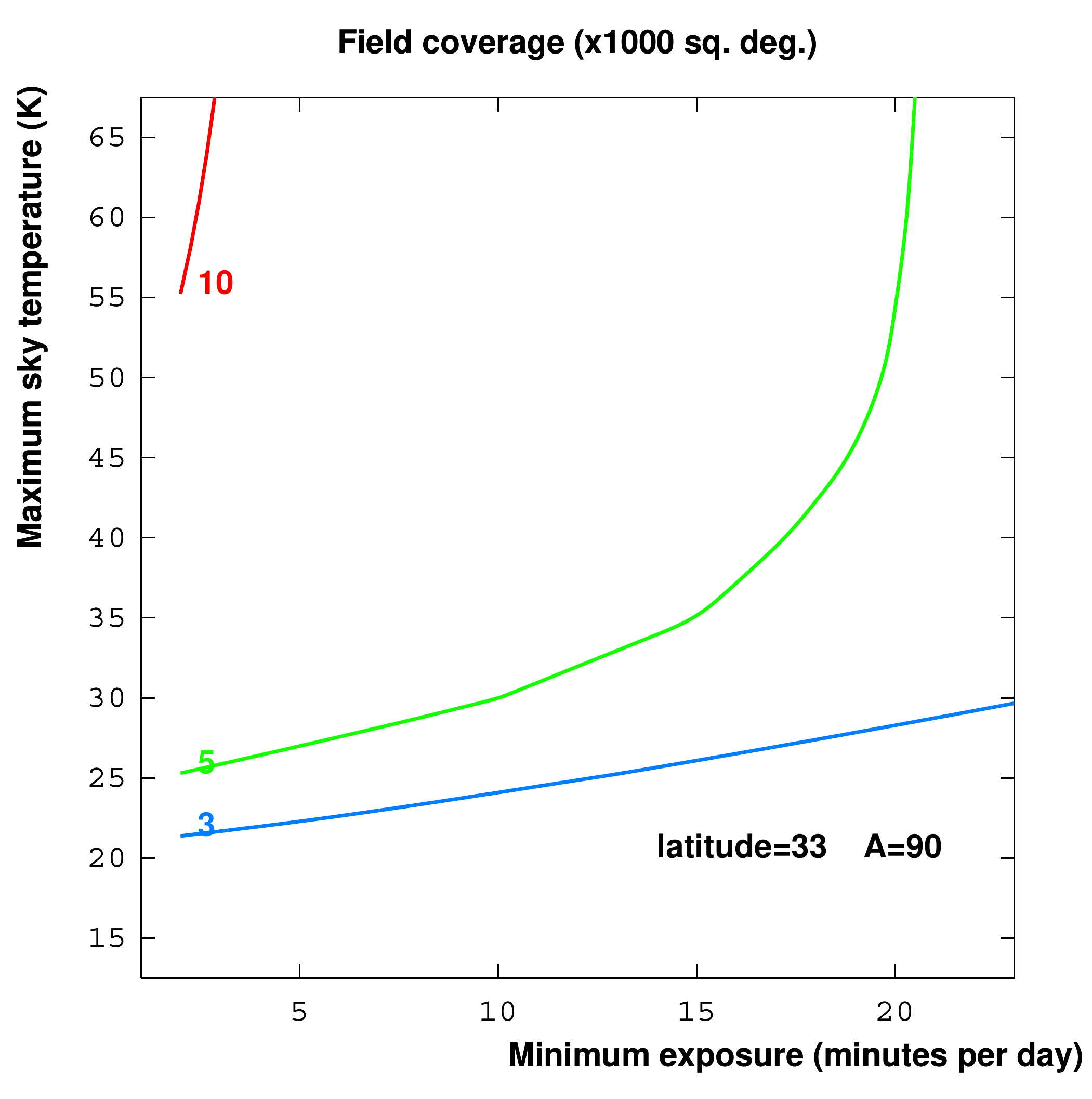}
\end{minipage}
\caption[]{
\it
Antenna located in central Marocco latitude.\\
LEFT:
Sky visibility (luminosity proportional to the Exposure time)
as a function of galactic coordinates.\\
From top to bottom:
antenna azimuth $A=0^\circ$ (North-South), $A=45^\circ$ and $A=90^\circ$ (East-West).\\
RIGHT:
field covered by the antenna as a function of the minimum daily exposure
and the maximum synchrotron sky temperature.
}
\label{maroc}
\end{figure}

Fig. \ref{maroc} (left) gives the exposure time for an antenna
with a $\Delta=2^\circ$ lobe, located
in central Morocco (latitude $33^\circ$) as a function of the
galactic coordinates for different orientations.
Fig. \ref{maroc} (right) gives the field covered by the antenna
with a daily exposure exceeding the abscissa-value and a sky synchrotron
temperature lower than the ordinate-value.
\begin{itemize}
\item
For $A=0^\circ$, 28200 square degree (68\%) of the sky are covered with a daily
exposure larger than 300s, and 1150 square degree (3\%) are covered with an
exposure larger than 1500s.
\item
For $A=45^\circ$, 22200 square degree (54\%) of the sky are covered with a daily
exposure larger than 300s, and 2100 square degree (5\%) are covered with an
exposure larger than 1500s.
\item
For $A=90^\circ$, 9600 square degree (23\%) of the sky are covered with a daily
exposure larger than 300s, and 4200 square degree (10\%) are covered with an
exposure larger than 1500s.
\end{itemize}
\subsection{South Africa}
\begin{figure}[h]
\centering
\begin{minipage}{7.0cm}
\includegraphics[width=6.5cm]{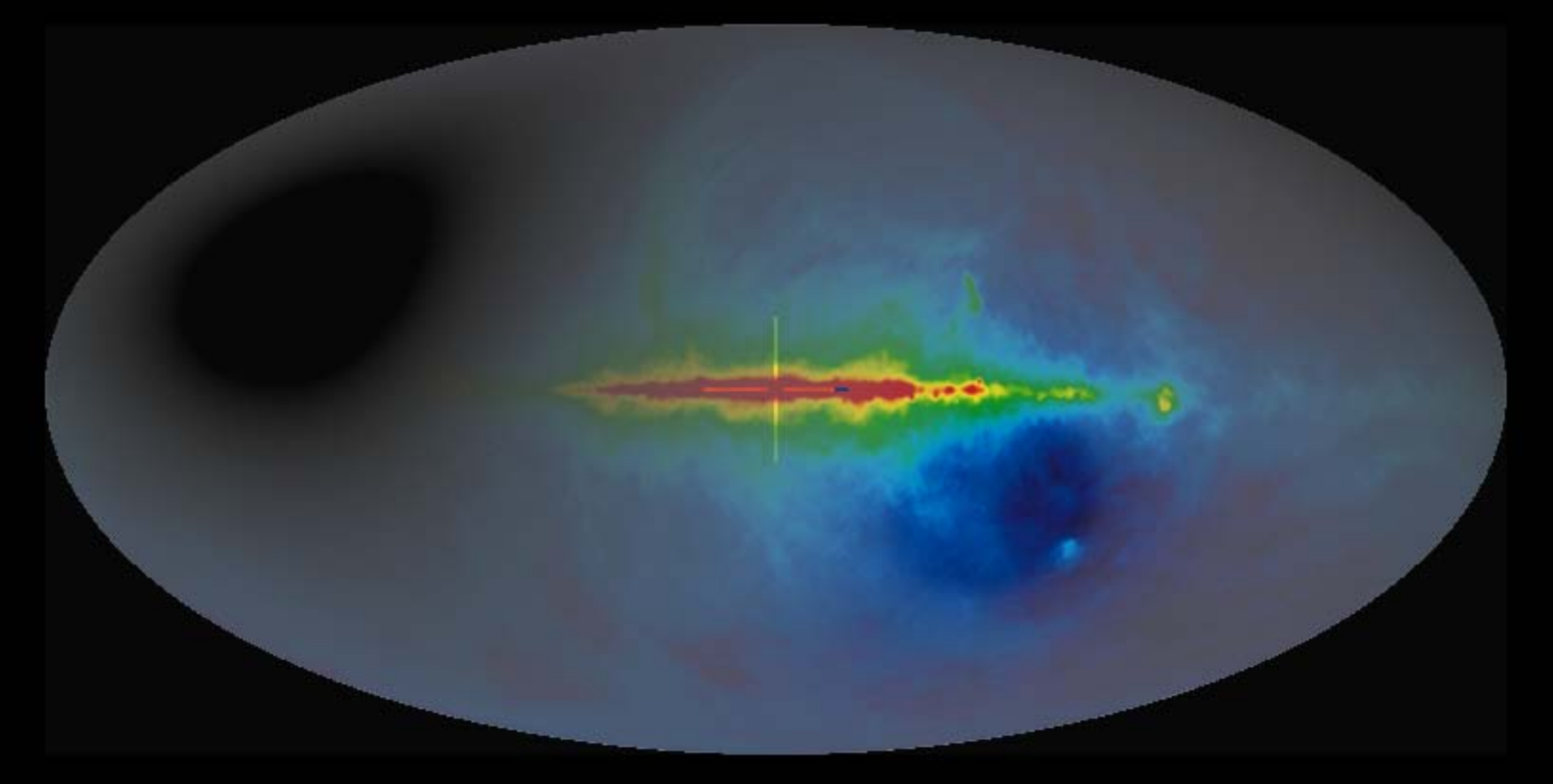}

\includegraphics[width=6.5cm]{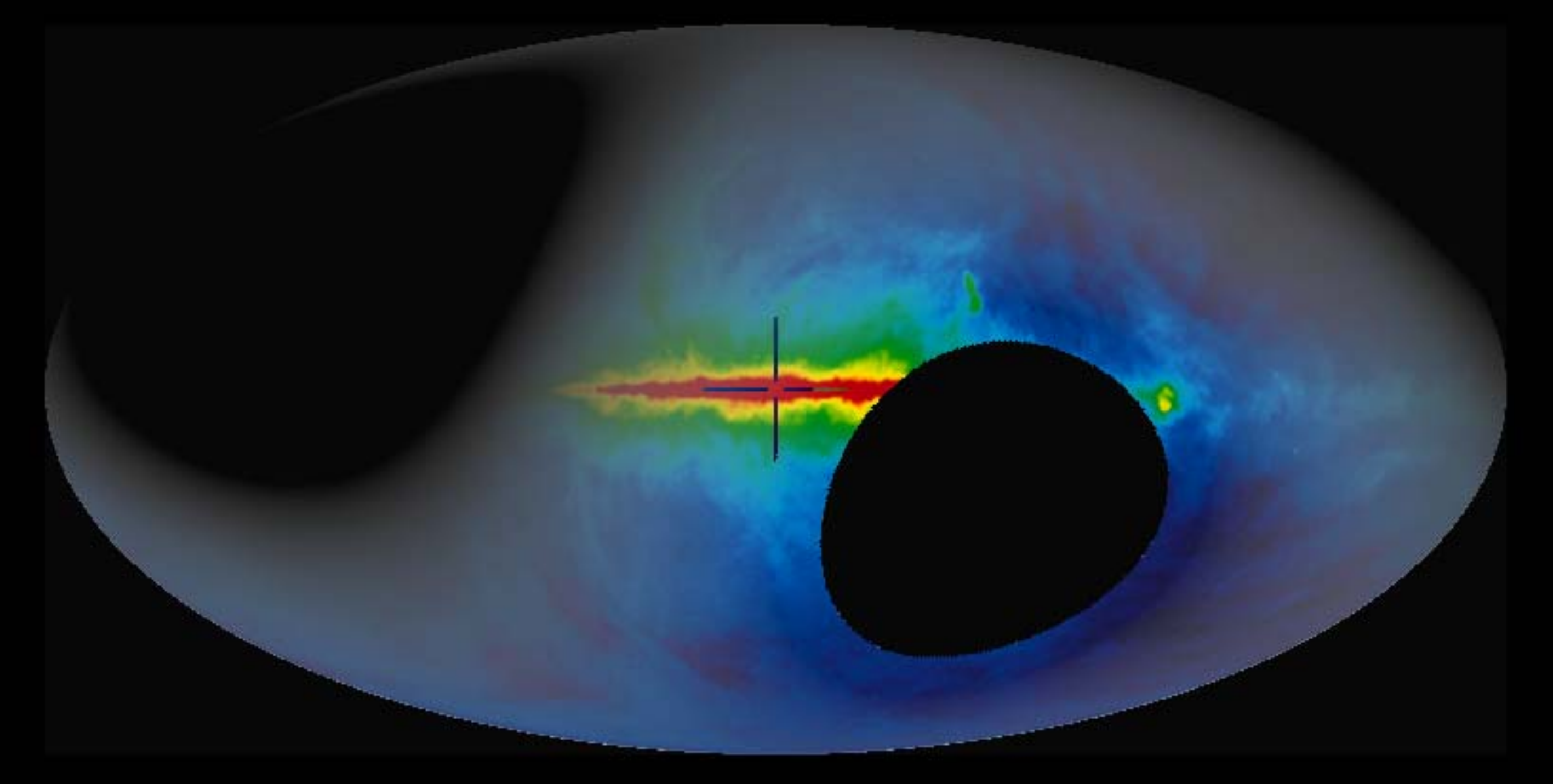}

\includegraphics[width=6.5cm]{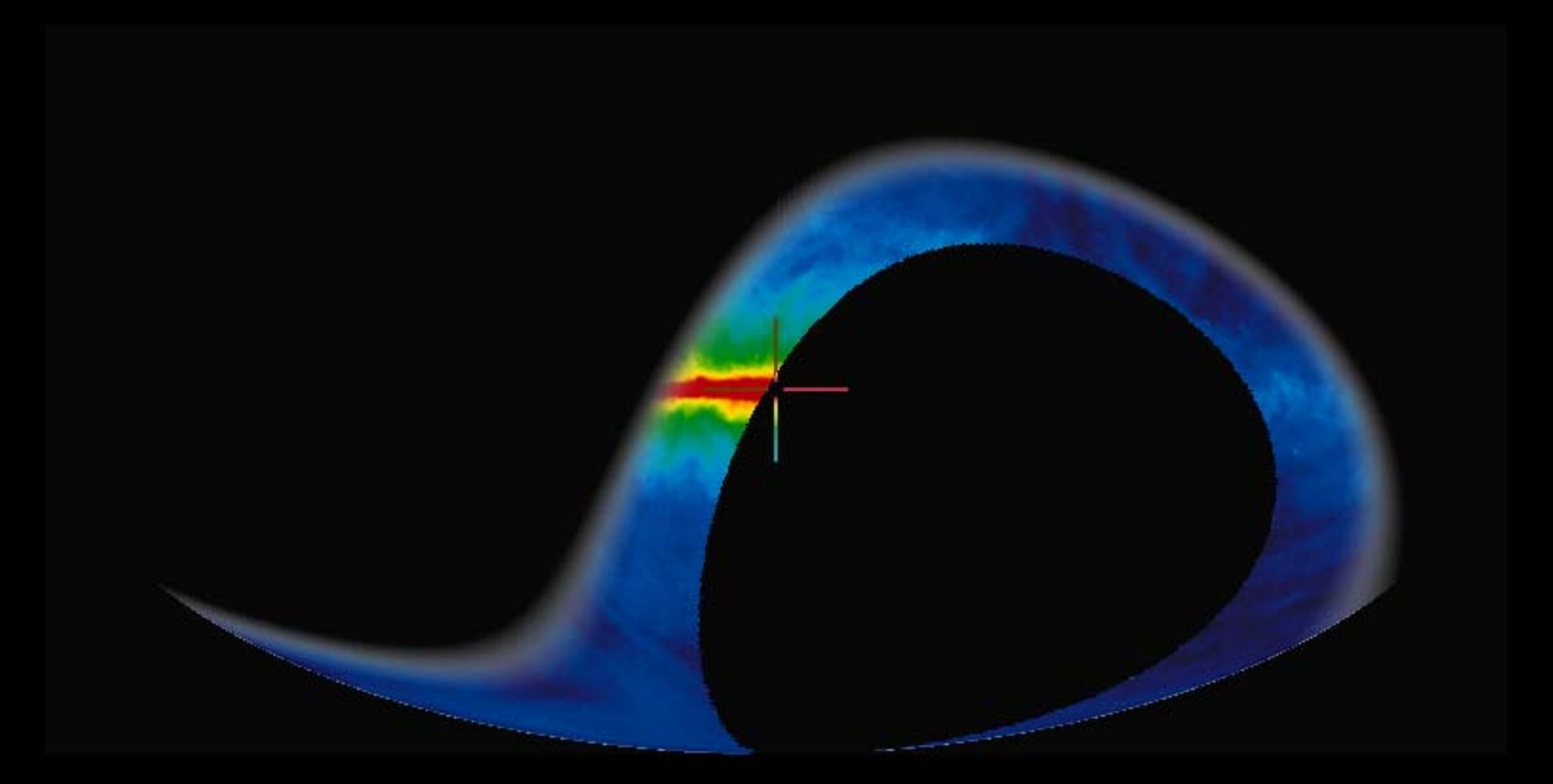}
\end{minipage}
\begin{minipage}{7.5cm}
\includegraphics[width=6.5cm,height=4.5cm]{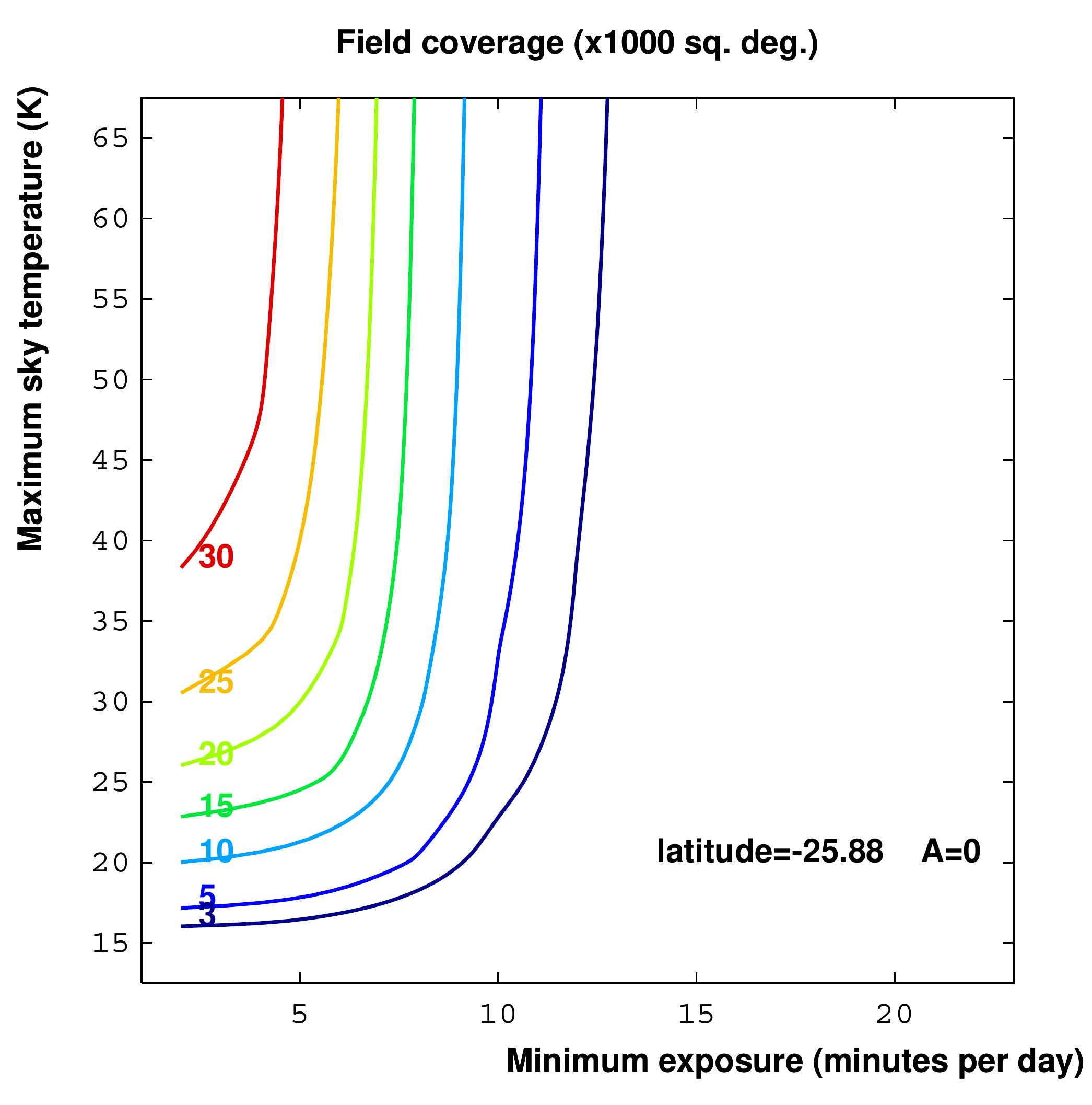}

\includegraphics[width=6.5cm,height=4.5cm]{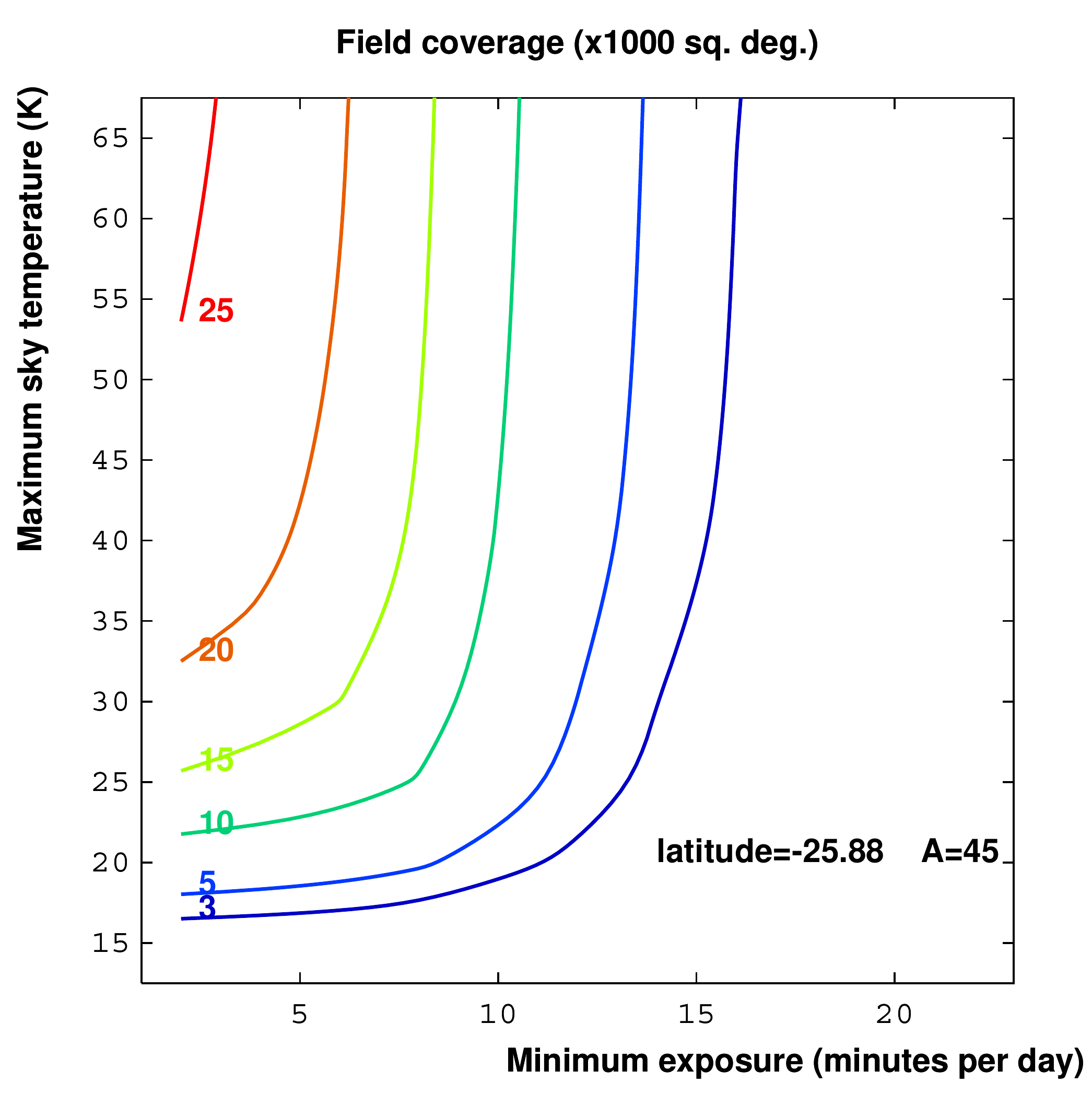}

\includegraphics[width=6.5cm,height=4.5cm]{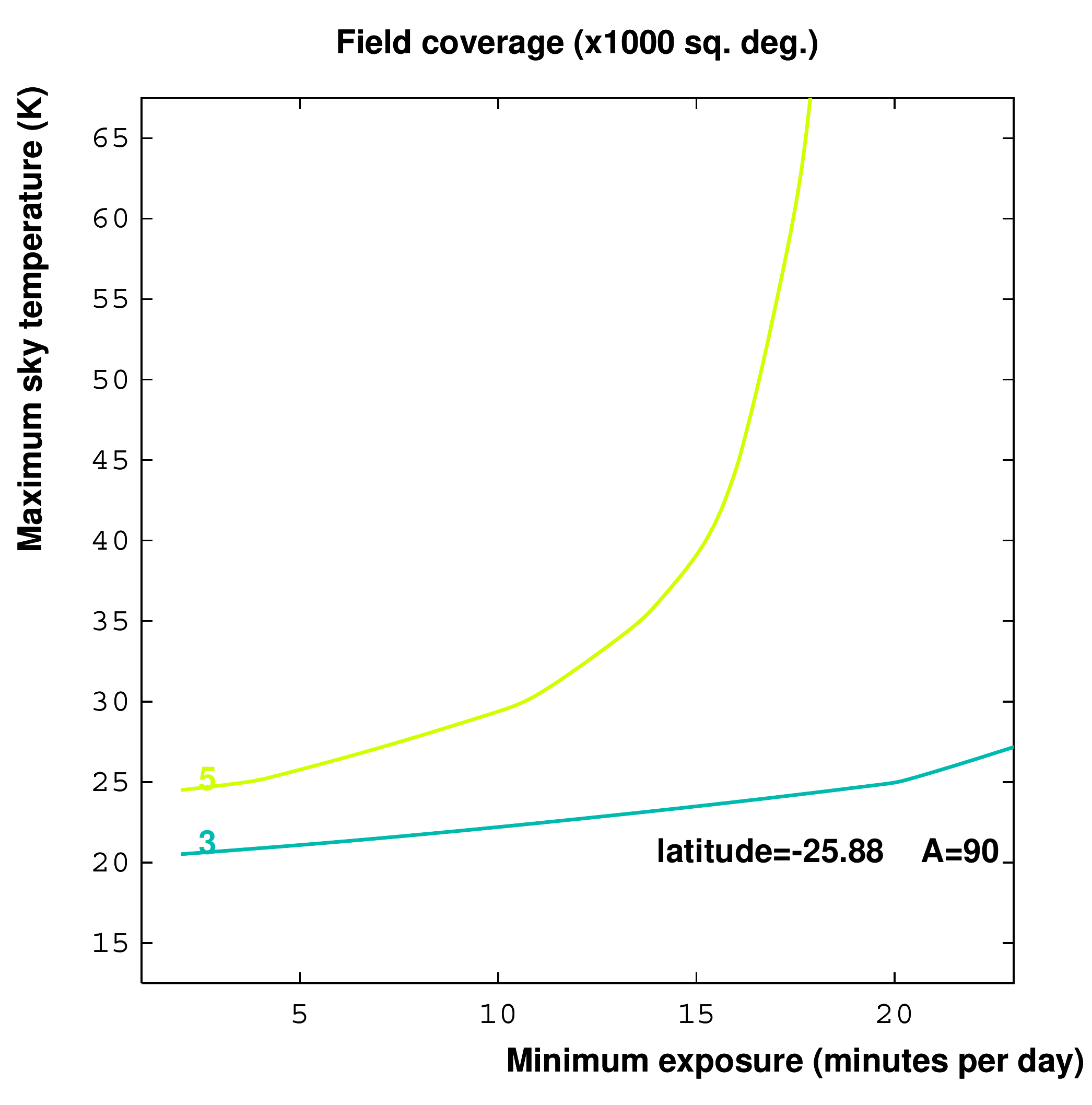}
\end{minipage}
\caption[]{ \it
Antenna located at Hartebeesthoek Radio Astronomy Observatory
latitude (South-Africa).\\
LEFT:
Sky visibility (luminosity proportional to the Exposure time)
as a function of galactic coordinates.\\
From top to bottom:
antenna azimuth $A=0^\circ$ (North-South), $A=45^\circ$ and $A=90^\circ$ (East-West).\\
RIGHT:
field covered by the antenna as a function of the minimum daily exposure
and the maximum synchrotron sky temperature.
}
\label{hartrao}
\end{figure}

Fig. \ref{hartrao} shows the exposure time and the field coverage
for an antenna located
at Hartebeesthoek Radio Astronomy Observatory (South Africa).

\begin{itemize}
\item
For $A=0^\circ$, 31600 square degree (77\%) of the sky are covered with a daily
exposure larger than 300s, and 760 square degree (2\%) are covered with an
exposure larger than 1500s.
\item
For $A=45^\circ$, 24300 square degree (59\%) of the sky are covered with a daily
exposure larger than 300s, and 1600 square degree (4\%) are covered with an
exposure larger than 1500s.
\item
For $A=90^\circ$, 8100 square degree (20\%) of the sky are covered with a daily
exposure larger than 300s, and 4300 square degree (10\%) are covered with an
exposure larger than 1500s.
\end{itemize}
\subsection{Equator}
\begin{figure}[h]
\centering
%
\begin{minipage}{7.5cm}
\includegraphics[width=7.cm]{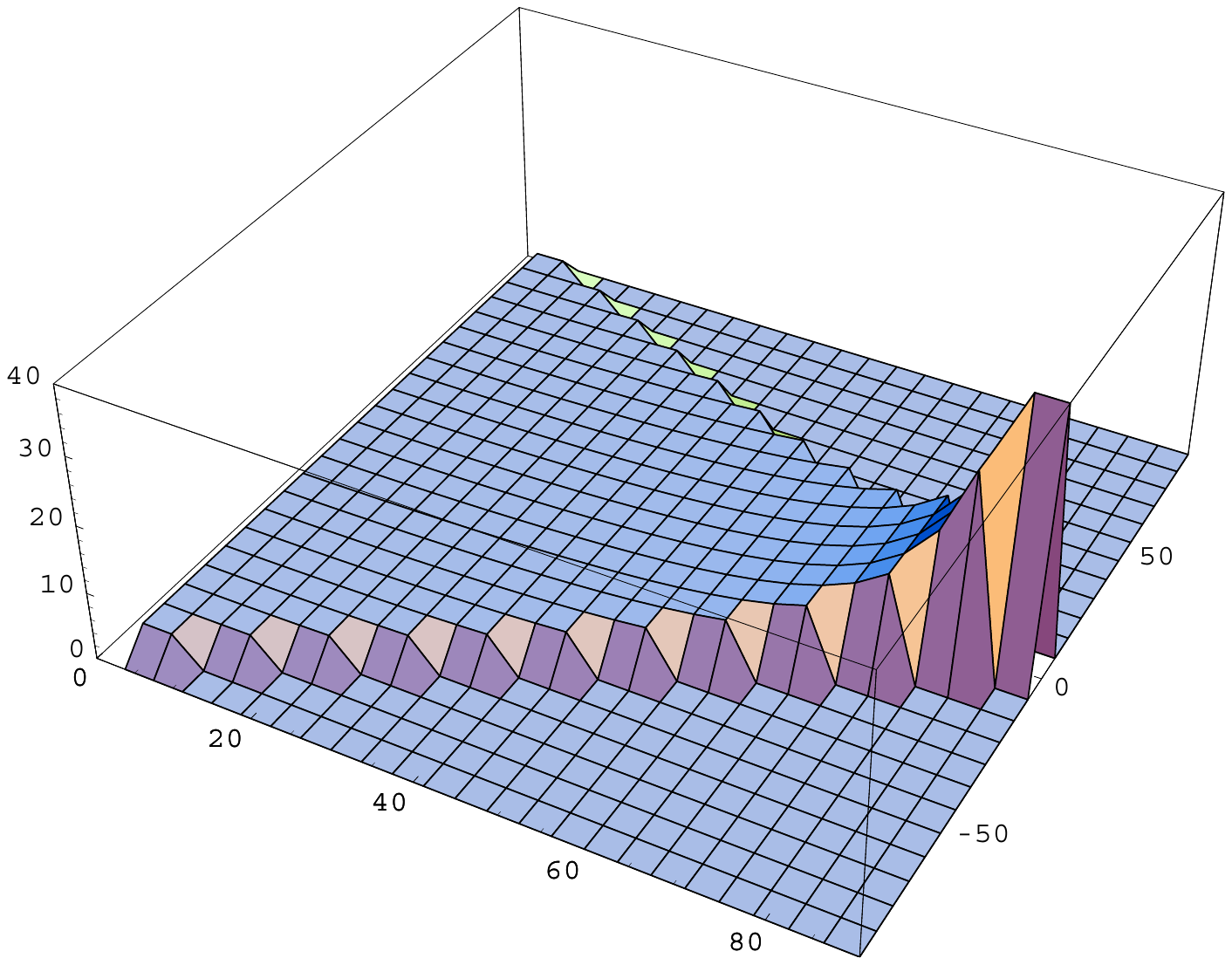}
\end{minipage}
\begin{minipage}{7.5cm}
\includegraphics[width=6.5cm]{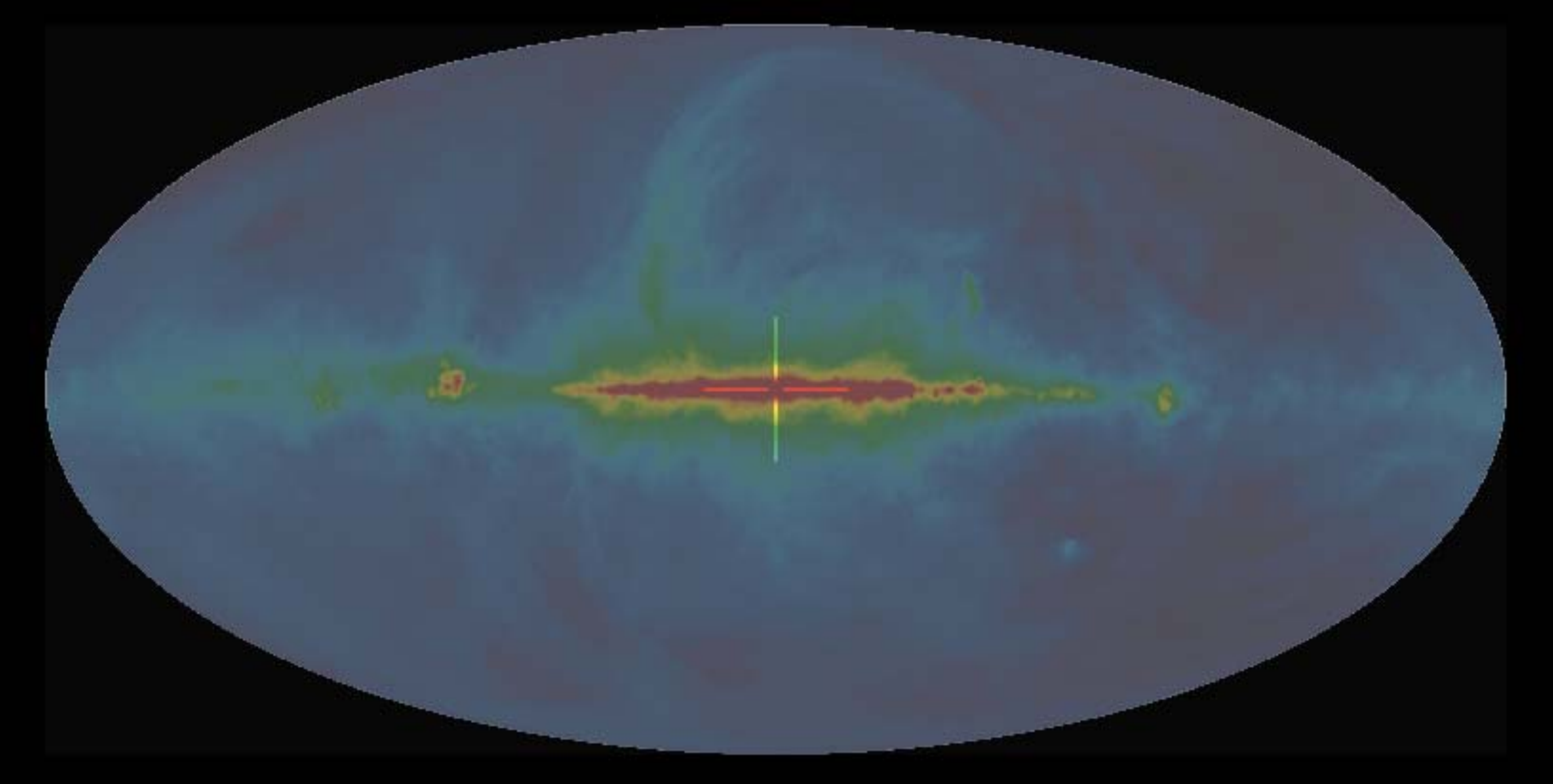}

\includegraphics[width=6.5cm]{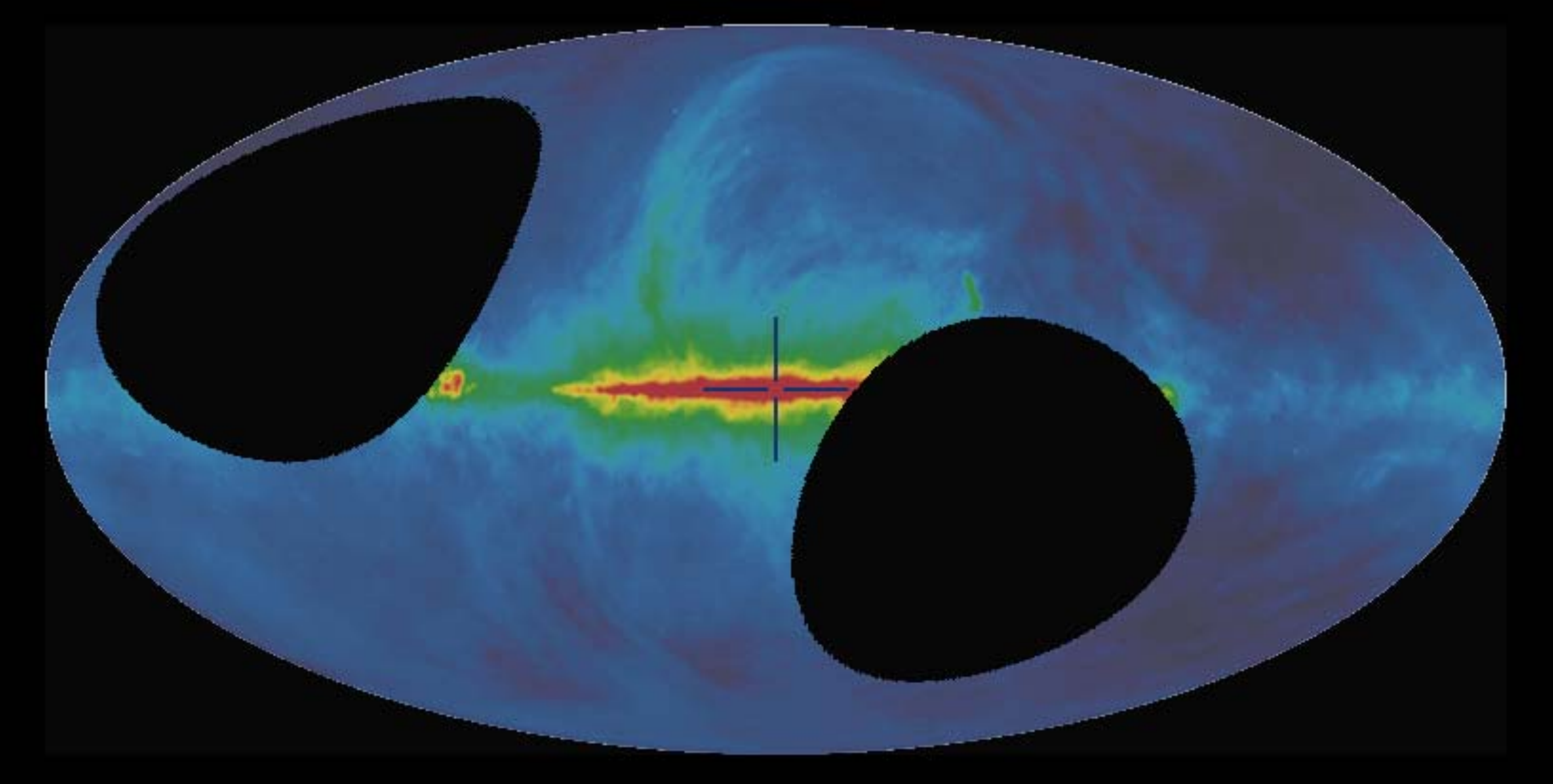}
\end{minipage}
\caption[]{
\it
LEFT: the exposure time as a function of the antenna azimuth $A$
(from 0 to $90^\circ$) and the declination (from $-90^\circ$
to $90^\circ$) in
the particular case of the equatorial location ($\lambda=0^\circ$).

RIGHT: exposure time as a function of galactic coordinates when the antenna has
azimuth $A=0^\circ$ (North-South, up) and $A=45^\circ$ (down).\\
}
\label{equator}
\end{figure}

Fig. \ref{equator} shows the exposure time at the
equator. In this case, the exposure time is uniform within
the complete observable field, whatever be the azimuth of the
antenna (but the observable field varies with $A$, see Fig. \ref{equator}).
For $A=0^\circ$, the daily exposure time is 480s on the full sky.
For $A=45^\circ$, 29000 square degree (71\%) of the sky are covered
with a daily exposure time of 679s.

\section{Conclusions}
The purpose of this study is to provide the exact expression
of the transit time of a given celectial object within the lobe of a
cylindrical reflector. Each particular case has been examined and
the results have been used to analyse different telescope configurations.
It is clear from this study that the groups planning to use a static
setup of cylinders should seriously consider the orientation
as a degree of freedom to favour either the largest field coverage
with the shortest mean transit time (North-South orientation),
or a smaller field coverage, but allowing a deeper survey (East-West
orientation).

\end{document}